\documentclass[
aps,
prd,
reprint,
twocolumn,
10pt,
nofootinbib,
amsfonts,
amssymb,
amsmath,
superscriptaddress,
floatfix]{revtex4}
\usepackage[colorlinks=true,allcolors=blue]{hyperref}
\usepackage{bm}
\usepackage{mathrsfs}
\usepackage[dvipsnames]{xcolor}
\usepackage{mathtools}
\usepackage{graphicx}
\usepackage{orcidlink}
\usepackage{placeins}

\newcommand{\AEI}{\affiliation{Max Planck Institute for Gravitational Physics (Albert Einstein Institute), Am M\"uhlenberg 1, D-14476 Potsdam, Germany}}
\newcommand{\Caltech}{\affiliation{Theoretical Astrophysics, Walter Burke Institute for Theoretical Physics, California Institute of Technology, Pasadena, California 91125, USA}}

\begin{document}

\author{Peter James Nee\orcidlink{0000-0002-2362-5420}}
\email{peter.nee@aei.mpg.de} \AEI
\author{Guillermo Lara \orcidlink{0000-0001-9461-6292}} \email{glara@aei.mpg.de} \AEI
\author{Harald P. Pfeiffer \orcidlink{0000-0001-9288-519X}} \AEI
\author{Nils L. Vu \orcidlink{0000-0002-5767-3949}} \Caltech

\date{\today}

\title{Quasistationary hair for binary black hole initial data
    \\ in scalar Gauss-Bonnet gravity}

\begin{abstract}
    Recent efforts to numerically simulate compact objects in alternative theories of gravity have largely focused on the time-evolution equations.
    Another critical aspect is the construction of constraint-satisfying initial data with precise control over the properties of the systems under consideration.
    Here, we augment the extended conformal thin sandwich framework to construct quasistationary initial data for black hole systems in scalar Gauss-Bonnet theory
    and numerically implement it in the open-source \texttt{SpECTRE} code.
    Despite the resulting elliptic system being singular at black hole horizons, we demonstrate how to construct numerical solutions that extend smoothly across the horizon.
    We obtain quasistationary scalar hair configurations in the test-field limit for black holes with linear/angular momentum as well as for black hole binaries.
    For isolated black holes, we explicitly show that the scalar profile obtained is stationary by evolving the system in time and compare against previous formulations of scalar Gauss-Bonnet initial data.
    In the case of the binary, we find that the scalar hair near the black holes can be markedly altered by the presence of the other black hole. The initial data constructed here enable targeted simulations in scalar Gauss-Bonnet simulations with reduced initial transients.

\end{abstract}

\maketitle


\section{Introduction} \label{sec: introduction}

Since the first gravitational wave (GW) event from a binary black hole coalescence, GW150914~\cite{LIGOScientific:2016aoc},
the possibility of testing our current theories of gravity against observational GW data in the highly dynamical strong-field regime has become a reality.
To date, while general relativity (GR) has been found to be consistent with current observations~\cite{LIGOScientific:2019fpa, LIGOScientific:2020tif, LIGOScientific:2021sio, EventHorizonTelescope:2021dqv, EventHorizonTelescope:2022xqj},
strong field tests for theories beyond GR have not yet been as thorough.
In the context of GWs, this is mostly due to the substantial effort
required to compute the detailed predictions needed to construct complete waveform models encompassing all stages of compact binary coalescence.
Crucially, accurate modeling of the highly nonlinear late-inspiral and merger stages relies on the ability to perform large-scale numerical relativity (NR) simulations~\cite{Baumgarte:2010ndz}.

In recent years, there has been growing interest in extending the techniques of NR to alternative theories of gravity.
Such theories are often motivated by open issues in gravity and cosmology, e.g., to provide a dynamical explanation to the observed accelerated expansion of the Universe or to connect GR to a more fundamental theory of quantum gravity.
For scalar tensor theories with two propagating tensor modes and one scalar mode~\cite{Horndeski:1974wa, Weinberg:2008hq, Langlois:2015cwa, Crisostomi:2016czh, BenAchour:2016fzp}, interactions between the metric and a dynamical scalar may lead to significant differences in the phenomenology of compact binaries.
For instance, in scalar Gauss-Bonnet gravity (sGB), the component black holes (BHs) in the binary may be endowed with \textit{scalar hair}~\cite{Sotiriou:2013qea, Sotiriou:2014pfa} and energy may be dissipated through radiation channels in addition to the two GW polarizations of  GR.

\begin{figure*}
    \includegraphics[width=0.49\linewidth]{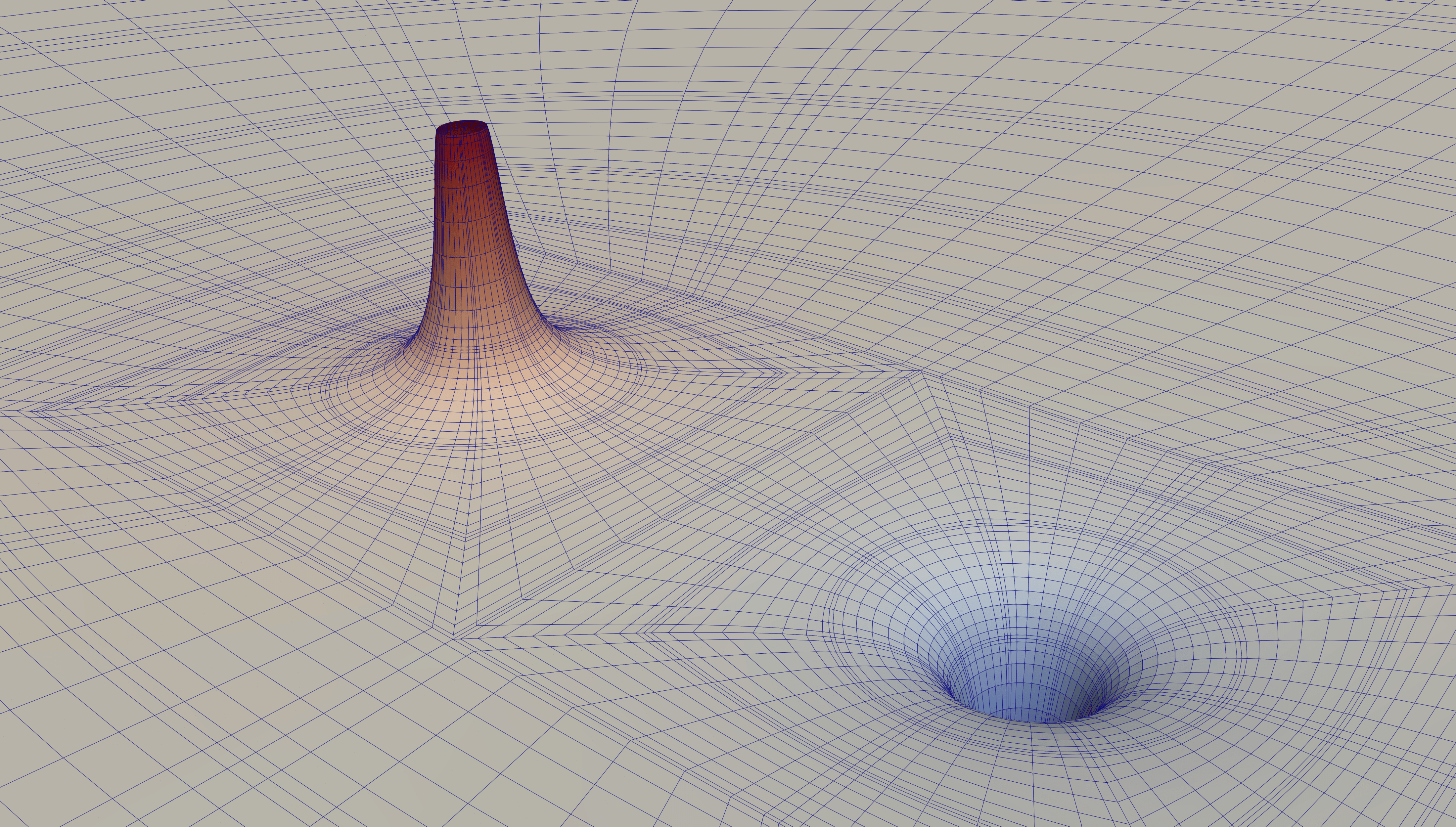}
    \includegraphics[width=0.49\linewidth]{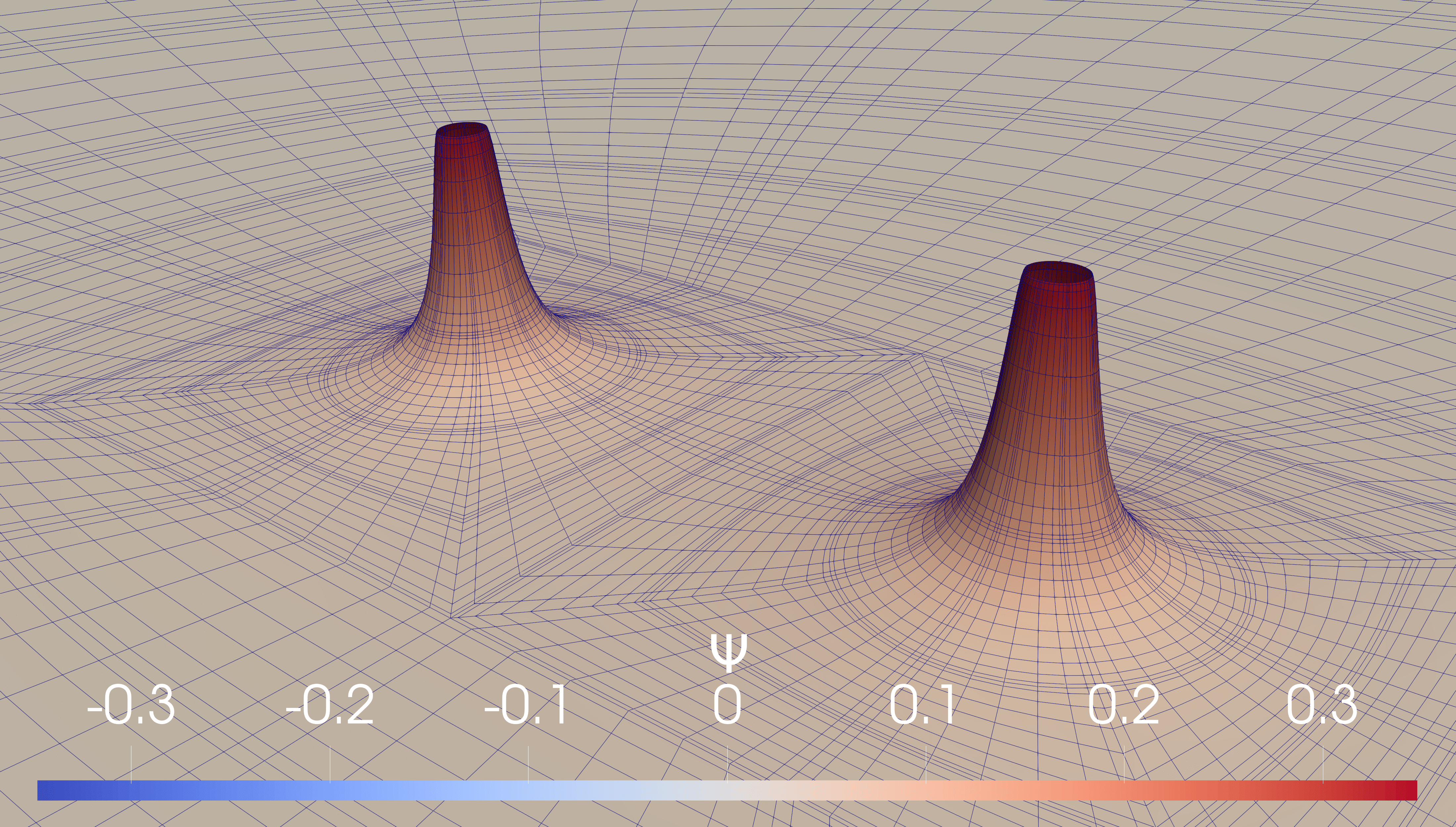}
    \caption{\textit{Scalar hair around a black hole binary.}
        \textbf{Left}: Opposite sign scalar charges. \textbf{Right}: Scalar charges with same sign.
        This configuration corresponds to an equal mass, nonspinning binary black hole system in an approximate circular orbit.
        The sGB coupling constants are $\ell^2 \eta \simeq 3.34M^2$, $\ell^2 \zeta \simeq -31.1M^2$, where $M$ denotes the mass of either BH, and  the
        BH separation  is $D  \simeq 28M$.
    }
    \label{fig: binary}
\end{figure*}

As is the case for GR, one can attempt to split the field equations in alternative theories of gravity into two sets of partial differential equations: a set of hyperbolic evolution equations, such as the generalized harmonic equations in GR; and a set of elliptic constraint equations, such as the Hamiltonian and momentum constraints in GR.
Nevertheless, the mathematical structure of both sets of equations differs from GR as the additional interactions contribute to new terms in the principal part.
In the strong field regime, such new terms may alter the hyperbolic/elliptic character of the equations and even lead to equations of mixed type [see, e.g.,~Refs.~\cite{Ripley:2019hxt, Bernard:2019fjb}].
In this respect, numerical relativity efforts have thus far focused on finding appropriate formulations for the set of evolution equations that allow for stable numerical evolutions.
These newly developed evolution strategies, which include novel gauges~\cite{Kovacs:2020PRL}, traditional perturbation theory techniques and proposals based on viscous hydrodynamics~\cite{Cayuso:2017iqc}
and their numerical implementation, have already produced a number of successful merger simulations in alternative gravity theories [see, e.g., Refs.~\cite{Okounkova:2019dfo, Okounkova:2019zjf, Okounkova:2020rqw, East:2020hgw, East:2021bqk, Figueras:2021abd, Bezares:2021dma, Corman:2022xqg, AresteSalo:2022hua, AresteSalo:2023mmd,  AresteSalo:2023hcp, Cayuso:2023aht, Held:2023aap, Kuan:2023trn, Corman:2024cdr}; Refs.~\cite{Okounkova:2017yby, Silva:2020omi, Elley:2022ept} for studies in the test-field approximation; and the review~\cite{Ripley:2022cdh}].

In this work, we take a step back and focus on the set of elliptic constraint equations.
Many of the current simulations for compact binary objects in scalar tensor theories either start off from initial data constructed for GR or use a superposition of isolated solutions.
While such approaches are practical and useful for first qualitative explorations, they are not guaranteed to satisfy the full constraint equations of the extended theory and will in general not be in quasistationary equilibrium.
Indeed, constraint-satisfying solutions can be obtained after an initial transient stage by employing standard techniques, e.g., by including constraint-damping terms or by smoothly turning on the additional interactions.
The cost, however, is a loss in control of the initial physical parameters (e.g., mass, spin, eccentricity)
during the relaxation stage
(which may migrate to different values), as well as the  additional computational resources spent in simulating this phase.
If our aim is to efficiently obtain accurate waveforms and to adequately cover the parameter space for the calibration of waveform models,
experience with GR has shown that constructing constraint-satisfying initial data in quasistationary equilibrium is important.

In GR, the most common way of formulating the Hamiltonian and momentum constraints as a set of elliptic equations is the conformal method, where instead of solving for geometric quantities directly one performs a conformal decomposition~\cite{Cook:2000vr}. This is the basis for two of the most well-known approaches, namely, the conformal transverse traceless (CTT)~\cite{OMurchadha:1974mtn} and the extended conformal thin sandwich (XCTS) methods~\cite{York:1998hy,Pfeiffer:2002iy}.

For the case of alternative theories, Kovacs~\cite{Kovacs:2021lgk} has recently examined the mathematical properties of the elliptic  systems arising in weakly coupled four-derivative scalar tensor  theories (a class of theories which includes the sGB theory investigated here) and provides theorems regarding the well-posedness of the boundary value problem using extensions of the CTT and XCTS methods.
On the practical side, several authors have constructed constraint-satisfying initial data for compact binaries in theories beyond GR.
Considering four-derivative scalar tensor theory, Ref.~\cite{Brady:2023dgu} prescribes an \textit{ad hoc} scalar field configuration, solving the constraint equations via a modification of the CTT approach~\cite{Aurrekoetxea:2022mpw}, in which the elliptic equation for the conformal factor is reinterpreted as an algebraic one for the mean curvature.
While the initial data constructed in this way is constraint-satisfying, since the scalar hair configuration is not in quasistationary equilibrium,
it should be expected to lead to significant transients during the initial stage of evolution.
A similar numerical approach is taken in Refs.~\cite{Siemonsen:2023age, Atteneder:2023pge} to obtain constraint satisfying initial data for boson star binaries, where the constraints are solved for free data specified by the superposition of isolated boson stars.
In the context of Damour-Esposito-Far\`ese
theory~\cite{Damour:1996ke} for neutron star binaries, Ref.~\cite{Kuan:2023hrh} solved the constraints for the metric alongside an additional Poisson equation for the scalar field.

This paper develops and implements a method to construct constraint-satisfying initial data where the scalar field is \textit{in equilibrium}.
We focus on the decoupling limit (i.e., the scalar does not back-react onto the metric) of scalar Gauss-Bonnet gravity in vacuum
\begin{align} \label{eq: action sGB}
    S\left[g_{ab}, \Psi \right] \equiv \int \, & d^4 x \sqrt{-g} \Big[\dfrac{R}{2 \kappa }
        - \dfrac{1}{2}  \nabla_{a} \Psi \nabla^{a} \Psi + \ell^2 f(\Psi) \, \mathcal{G}\Big],
\end{align}
where \(\kappa \equiv 8\pi G\), \(\ell\) denotes the coupling constant, \(g = \mathrm{det}(g_{ab})\) is the determinant of the metric \(g_{ab}\), and
\(\Psi\) is the scalar field.
To obtain spontaneously scalarized BHs~\cite{Silva:2017uqg, Doneva:2017bvd, Antoniou:2017acq} we choose the free function \(f(\Psi)\) as~\cite{Silva:2018qhn}
\begin{align} \label{eq: coupling function quartic model}
    f(\Psi) \equiv \dfrac{\eta}{8} \Psi^2 + \dfrac{\zeta}{16} \Psi^4.
\end{align}
This function couples $\Psi$ to the Gauss-Bonnet scalar
\begin{align}
    \mathcal{G} \equiv R_{abcd}R^{abcd} - 4 R_{ab}R^{ab} + R^2,
\end{align}
which is in turn defined in terms of the Riemann tensor \(R_{abcd}\), the Ricci tensor \(R_{ab}\), and the Ricci scalar \(R\).

Following Ref.~\cite{Kovacs:2021lgk}, we revisit the conditions for obtaining quasistationary configurations for the scalar hair around isolated and binary black holes.
We argue that, in the initial data slice, one must impose a vanishing scalar ``momentum'' defined in terms of the directional derivative along an approximate Killing vector of the spacetime--as opposed to the directional derivative along the normal to the foliation as in Ref.~\cite{Kovacs:2021lgk}.
The adapted coordinates from the background spacetime given by a solution to the XCTS equations naturally yield the required approximate Killing vector.
Imposing the appropriate momentum condition on the scalar equation, we derive a \textit{singular} boundary-value problem for BH spacetimes.

We demonstrate that this singular boundary-value problem can be solved without an inner boundary condition in the spectral elliptic solver of the open-source \texttt{SpECTRE} code~\cite{deppe_2024_11494680}.
We thus obtain quasistationary hair for both single and binary black hole spacetimes, as illustrated in Fig.~\ref{fig: binary}.
Moreover, for the case of single BHs, we further evolve the obtained configurations to confirm that the solution is indeed quasistationary and does not lead to large transients, and
compare against the prescription given in Ref.~\cite{Kovacs:2021lgk}.

This paper is organized as follows.
Section~\ref{sec: theory} recalls basic aspects of sGB theory and of the XCTS method.
In Sec.~\ref{sec: single black holes}, we revisit different formulations  for the scalar equation and define a scheme that imposes quasiequilibrium on the scalar hair.
We further discuss the singular boundary value problem and describe our numerical implementation to solve for single BHs. Section \ref{sec: single black holes results} showcases the different single BH systems that we can access with both the 1D and 3D code, and demonstrates the stationarity of the solutions obtained via direct evolution.

Section \ref{sec: binary black holes} constructs initial data for binary black holes with scalar hair. We first deal with conceptual issues regarding scalar configuration on arbitrarily large spatial domains and then proceed to present our solutions for quasistationary scalar hair.
We summarize and discuss our results in Sec.~\ref{sec: conclusion}.
Throughout this paper, we use geometric units such that \(G = c = 1\) and \((-+++)\) signature.
Early alphabet letters \(\{a,b,c\}\) represent 4-dimensional spacetime indices, while middle alphabet letters \(\{i,j,k\}\) correspond to 3-dimensional spatial indices.


\section{Theory} \label{sec: theory}

Variation of the action of scalar Gauss-Bonnet theory [Eq.~\eqref{eq: action sGB}] yields a scalar equation
\begin{align} \label{eq: scalar equation sGB}
    \Box \Psi & = - \ell^2 f'(\Psi) \mathcal{G}
\end{align}
and a tensor equation
\begin{align} \label{eq: tensor equation}
    R_{ab} & = H_{ab}\left[g_{cd}, \Psi\right],
\end{align}
where \(H_{ab}\left[g_{cd}, \Psi\right]\) contains up to second derivatives of \(g_{ab}\) and \(\Psi\) --see, e.g., Ref.~\cite{Franchini:2022ukz} for the full expression.
In the decoupling limit of the theory (i.e.~when \(\Psi\) is considered a test field), the right-hand-side of Eq.~\eqref{eq: tensor equation} vanishes, \(H_{ab}\left[g_{cd}, \Psi\right] \equiv 0\).


\subsection{Spontaneous scalarization}


Stationary BH solutions of Eqs.~\eqref{eq: scalar equation sGB} and \eqref{eq: tensor equation} are often nonunique.
When $f'(0)=0$, as for our choice of $f(\Psi)$, a GR solution with $\Psi\equiv0$ trivially solves Eqs.~\eqref{eq: scalar equation sGB} and \eqref{eq: tensor equation}.
However, GR solutions can be energetically disfavored for a large enough coupling parameter \(\ell^2 \eta \gtrsim 0\).
This can be seen~\cite{Doneva:2022ewd} by expanding around \(\Psi \equiv 0\) to derive an equation describing the scalar perturbations around the GR solution,
\begin{align}\label{eq: perturbative}
    (\Box  - m^2_{\Psi, \mathrm{eff}} )\, \delta \Psi= 0,
\end{align}
where $m^2_{\Psi, \mathrm{eff}} \equiv -\ell^2 \eta \mathcal{G}$ plays the role of an effective, spatially varying mass term.
If \(m^2_{\Psi, \mathrm{eff}}\) is negative enough due to either high curvature or very large spins~\cite{Dima:2020yac, Doneva:2020nbb}, GR solutions in sGB may become dynamically unstable, and will \textit{spontaneously scalarize} to yield a second set of solutions with nonvanishing scalar \textit{hair}~\cite{Silva:2017uqg,Doneva:2017bvd,Antoniou:2017acq, Doneva:2023oww}.
Therefore, BHs in sGB theory are characterized by their mass, spin, and an additional \textit{scalar charge} parameter \(q\), defined by the asymptotic behavior of the scalar as
\begin{align} \label{eq: Psi asymptotic dacay}
    \Psi (r \to \infty ) = \Psi_\infty + \dfrac{q M^2}{r} + \mathcal{O}\left(\dfrac{1}{r^2}\right),
\end{align}
where \(\Psi_\infty\) is the asymptotic value of the scalar field and \(M\) is the mass of the BH.
Given the \(\Psi \to -\Psi\) symmetry of the theory described by Eqs.~\eqref{eq: action sGB} and \eqref{eq: coupling function quartic model}, any hairy solutions will have a corresponding equivalent solution related by \(\Psi \to -\Psi\), and which is characterized by a scalar charge of equal magnitude and opposite sign.


\subsection{The XCTS formulation}

In the decoupling limit, the constraint equations arising from Eq.~\eqref{eq: tensor equation} are the usual Hamiltonian and momentum constraints of GR.
To obtain them, we perform a (3+1)-decomposition of the metric,
\begin{align} \label{eq: 3+1 decomposition of the metric}
    ds^2
     & = g_{ab} \, dx^a \, dx^b \nonumber                                               \\
     & = - \alpha^2 \, dt^2 + \gamma_{ij} (\beta^i \, dt + dx^i)(\beta^j \, dt + dx^j),
\end{align}
where \(\alpha\) is the lapse, \(\beta^i = \gamma^{ij}\beta_{j}\) is the shift, and \(\gamma_{ij}\) is the spatial metric (with inverse \(\gamma^{ij}\)).
The constraints in vacuum read~\cite{Baumgarte:2010ndz}
\begin{subequations}
    \label{eq: Hamiltonian and momentum constraints}
    \begin{align}
        {^{(3)} R} + K^2 - K_{ij} K^{ij}             & = 0, \\
        D_{j} \left({K^{ij}} - \gamma^{ij} K \right) & =0,
    \end{align}
\end{subequations}
where \({^{(3)} R}\) denotes the Ricci scalar of $\gamma_{ij}$ and \(D_j\) is the 3-dimensional covariant derivative compatible with $\gamma_{ij}$.
Finally,
\(K_{ab} \equiv - (1/2)\mathcal{L}_{n} \gamma_{ab}\) denotes the extrinsic curvature, with trace $K$, where the Lie derivative is taken along the future-pointing unit normal to the foliation, $n^a$.

We  further decompose the spatial metric as
\begin{align}
    \gamma_{ij} & = \psi^4 \bar{\gamma}_{ij},
\end{align}
where \(\psi > 0\) is the conformal factor and \(\bar{\gamma}_{ij}\) is the conformal spatial metric, which we are free to specify. The XCTS formalism~\cite{York:1998hy,Pfeiffer:2002iy} is centered around
specifying certain free data and their \textit{time derivatives}.  Specifically, the conformal metric $\bar\gamma_{ij}$ and $K$ are free data, as well as
$\partial_t\bar\gamma_{ij}\equiv \bar u_{ij}$ and $\partial_tK$.  It is useful to decompose the extrinsic curvature as
\begin{equation}
    K^{ij} = \frac{1}{3}\gamma^{ij}K + \psi^{-10}\bar A^{ij}
\end{equation}
with
\begin{align}
    \bar{A}^{i j}=\frac{1}{2\bar\alpha}\left[(\bar{L} \beta)^{i j}-\bar{u}^{i j}\right].
\end{align}
Here, $\bar\alpha= \psi^{-6}\alpha$ is the conformal lapse function, and the conformal longitudinal operator is defined as
\begin{align}
    (\bar{L} \beta)^{i j}=2 \bar{D}^{(i} \beta^{j)}-\frac{2}{3} \bar{\gamma}^{i j} \bar{D}_k \beta^k,
\end{align}
where $\bar D_i$ denotes the covariant derivative operator compatible with the conformal metric $\bar\gamma_{ij}$.

The final XCTS equations are then obtained from Eqs.~\eqref{eq: Hamiltonian and momentum constraints} and from the evolution equation for \(K\), and are given by~\cite{Pfeiffer:2002iy}
\begin{subequations}\label{eq: XCTS equations}
    \begin{align}
        \bar{D}^2 \psi- \frac{1}{8} \psi\, {^{(3)}\!\bar{R}}-\frac{1}{12} \psi^5 K^2+\frac{1}{8} \psi^{-7} \bar{A}_{i j} \bar{A}^{i j} & =0,                                  \\
        \bar{D}_i\bigg(\frac{1}{\bar\alpha}\big[(\bar{L} \beta)^{i j}-\bar u^{ij}\big]\bigg)-\frac{2}{3} \psi^6 \bar{D}^j K            & =0,
        \\
        \bar{D}^2(\alpha \psi)- \alpha \psi\bigg(\frac{7}{8} \psi^{-8} \bar{A}_{i j} \bar{A}^{i j}+\frac{5}{12} \psi^4 K^2+            & \frac{1}{8} {^{(3)}\!\bar{R}}\bigg)  \\
        +\psi^5 \big(\partial_t K+                                                                                                     & \beta^i \bar{D}_i K\big)=0,\nonumber
    \end{align}
\end{subequations}
where \({^{(3)}\!\bar{R}}\) is the spatial conformal Ricci scalar.
In the XCTS formalism, the notion of quasistationary equilibrium can be imposed~\cite{Cook:2004kt} by demanding that the conformal metric and trace of the extrinsic curvature remain unchanged along infinitesimally separated spatial slices, i.e.
\begin{align}
    \bar u_{ij}  & = 0, \\
    \partial_t K & = 0.
\end{align}

Combined with appropriate boundary conditions [see Ref.~\cite{Cook:2004kt} for details], the XCTS system [Eqs.~\eqref{eq: XCTS equations}] is then solved for \(\{\psi, \alpha\psi, \beta^{i}\}\), thus providing not only a solution to the constraint equations \eqref{eq: Hamiltonian and momentum constraints}, but also a coordinate system adapted to the symmetry along the approximate Killing vector
\begin{align}
    t^{a} \partial_{a} = \left(\alpha n^{a} + \beta^{a}\right)\partial_{a} = \partial_t.
\end{align}
In Sec.~\ref{sec: single black holes}, we will extend this property to the scalar equation in sGB theory.


\section{Quasistationary scalar hair} \label{sec: single black holes}

In this section we revisit the scalar equation Eq.~\eqref{eq: scalar equation sGB} and consider different strategies to include it in the XCTS scheme.
The aim is to obtain solutions for the metric \textit{and} the scalar hair of the BH in a general 3-dimensional space without symmetry.
We further describe our numerical implementation,
which will also be applicable to the more general case of BH binaries treated in Sec.~\ref{sec: binary black holes}.


\subsection{Spherical symmetry}
\label{sec: theory solving 1d problem}

We first consider a spherically symmetric BH in horizon-penetrating Kerr-Schild coordinates
\begin{multline} \label{eq: KerrSchild metric spherical symmetry}
    ds^2 = -\left(1 - \dfrac{2M}{r} \right) dt^2 + \dfrac{4 M}{r} dt\, dr\\
    + \left(1 + \dfrac{2 M}{r}\right) dr^2 + r^2 \, d\Omega^2 ~,
\end{multline}
with \(d \Omega^2 = d \theta^2 + \sin^2(\theta) d \phi^2\). Under the assumption that the scalar field is time-independent, Eq.~\eqref{eq: scalar equation sGB} yields
\begin{align} \label{eq: scalar equation in KerrSchild}
    \left(1 - \dfrac{2M}{r} \right) \partial^2_r \Psi - \dfrac{2 (M-r)}{r^2} \partial_r \Psi  = -\dfrac{48 M^2}{r^6} f'(\Psi),
\end{align}
where \(\mathcal{G} = 48 M^2 / r^6\) and where \(M\) is the mass of the BH.
We will be looking for solutions of Eq.~\eqref{eq: scalar equation in KerrSchild} with asymptotic behavior~\eqref{eq: Psi asymptotic dacay} by imposing\footnote{
    While one can easily place the outer boundary $r_{\rm max}$ at spatial infinity in the spherically symmetric case, we impose the condition~\eqref{eq: decay boundary condition} to connect with the 3D-implementation in Sec.~\ref{sec: single BH 3D}.}
\begin{align} \label{eq: decay boundary condition}
    \left[r \partial_r \Psi + \Psi - \Psi_\infty \right]_{r \to \infty}= 0 ~,
\end{align}
with \(\Psi_\infty = 0\).

This is our first encounter with a \textit{singular} boundary value problem.
Notice that Eq.~\eqref{eq: scalar equation in KerrSchild} is singular at the BH horizon \(r_{h} =  2 M\), where the factor in front of the highest-derivative operator vanishes.
Despite this observation, Eq.~\eqref{eq: scalar equation in KerrSchild} can be easily solved via the shooting method~\cite{press2007numerical}. Regularity of \(\Psi\) at the horizon can be imposed by expanding $\Psi$ as an analytic series around \(r_h\). The solutions satisfying Eq.~\eqref{eq: decay boundary condition} can then be found by numerically integrating outward starting from \(r_h + \epsilon\) and performing a line search in the unknown value \(\Psi \rvert_{r_h}\) at the inner boundary.

In order to prepare for our later 3D solutions, we shall take a different approach, and solve Eq.~\eqref{eq: scalar equation in KerrSchild} on a domain that extends through the horizon into the interior region.  Utilizing a spectral method, we represent $\Psi$ as a series in Chebychev polynomials \(T_{i}(x)\),
\begin{align} \label{eq: Psi in Chebychev basis}
    \Psi(x) = \sum_{i = 0}^{N} \Psi_{(i)} T_{i}(x),
\end{align}
where the argument \(x \in [-1, 1]\) is related to radius \(r\) by the transformation \(x = A / (r - B) + C\) for suitable constants $A$, $B$, and $C$. To cover $r\in[r_{\rm min}, r_{\max}]$, we set $A=(r_{\rm min}+r_{\rm max}+2C)/(r_{\rm max}-r_{\rm min})$, $B=r_{\rm max}-Ar_{\rm max}+C-AC$, and leave $C$ as a specifiable constant to adjust the distribution of resolution throughout the interval.
We choose a spatial grid \(\{x_i\}_{i = 0}^{N}\) defined by the nodes (or zeros)  of \(T_{i}(x)\), and
compute spatial derivatives of \(\Psi\) analytically from Eq.~\eqref{eq: Psi in Chebychev basis}.
Using a Newton-Raphson scheme, we iteratively solve the scalar equation by expanding \(\Psi \to \Psi + \delta \Psi \) and linearizing Eq.~\eqref{eq: scalar equation in KerrSchild}. We obtain
\begin{multline} \label{eq: scalar equation iterative scheme spherical symmetry}
    \left(1 - \dfrac{2M}{r} \right) \partial^2_r \delta \Psi^{(K)} - \dfrac{2 (M-r)}{r^2} \partial_r \delta \Psi^{(K)} \\
    + \dfrac{48 M^2}{r^6} f''\left(\Psi^{(K)}\right) \delta \Psi^{(K)} \\
    = - \left(1 - \dfrac{2M}{r} \right) \partial^2_r \Psi^{(K)} + \dfrac{2 (M-r)}{r^2} \partial_r \Psi^{(K)} \\
    -\dfrac{48 M^2}{r^6} f'\left(\Psi^{(K)}\right),
\end{multline}
where, at a given iteration step \(K\), the improved solution is given by
\begin{align}
    \Psi^{(K+1)} = \Psi^{(K)} + \delta \Psi^{(K)}.
\end{align}

\begin{figure}
    \includegraphics[width=\linewidth]{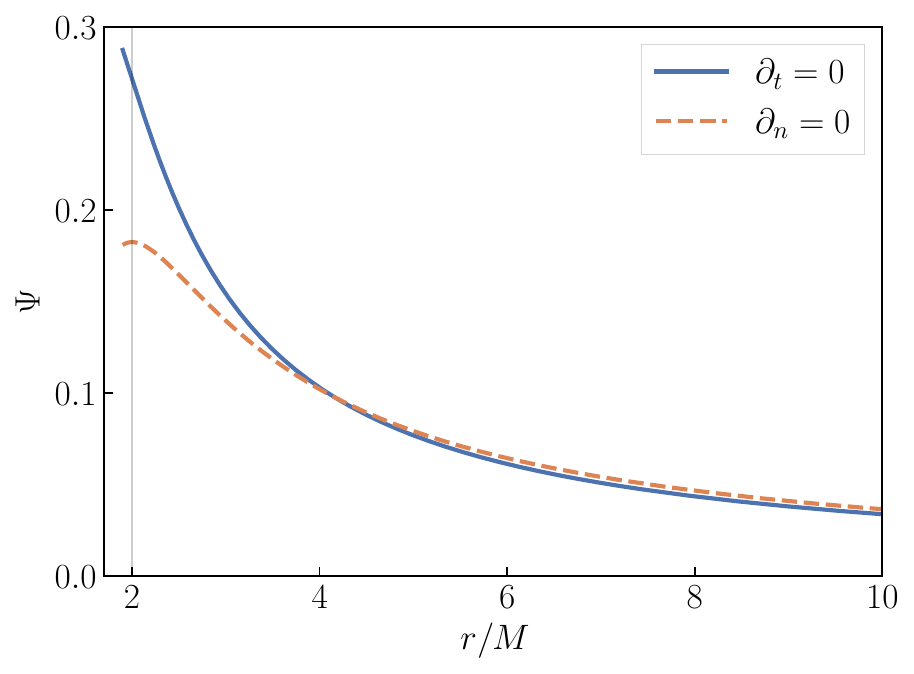}
    \caption{\label{fig: 1d profiles} \textit{Radial profile of the scalar field around a
        nonrotating BH.} The curves correspond to different formulations used for obtaining $\Psi$. We set
    $\ell^2\eta/M^{2}=6$, $\ell^2\zeta/M^{2}=-60$. The vertical gray line indicates the location of the horizon.}
\end{figure}

For a solution interval crossing the horizon, i.e., $r_{\rm min} < r_h <  r_{\rm max}$, we impose boundary conditions of the form~\eqref{eq: decay boundary condition} only at the outer boundary. We do \textit{not} impose regularity across the entire domain (in particular, at a singular boundary at \(r = r_h\)) via boundary conditions, as it is already built into the spectral expansion~\eqref{eq: Psi in Chebychev basis} since all Chebychev polynomials are regular. We implement this algorithm in \texttt{Python}, and for each iteration step $K$ we solve the discretized version of Eq.~\eqref{eq: scalar equation iterative scheme spherical symmetry} via explicit matrix inversion using \texttt{NumPy}. An exemplary solution of Eq.~\eqref{eq: scalar equation in KerrSchild} is shown as the blue line in Fig.~\ref{fig: 1d profiles}, where we set $r_{\rm min}=1.9M$ and $r_{\rm max}=10^{10}M$.

\subsection{\boldmath 3D normal formulation ``$\partial_n=0$''}

To solve for scalar hair in a general 3-dimensional space, Ref.~\cite{Kovacs:2021lgk} requires the ``momentum''
\begin{align}\label{eq: momentum Pi definition}
    \Pi \equiv - n^{a} \partial_{a} \Psi
\end{align}
to vanish everywhere on the initial spatial slice at \(t = 0\), i.e., \(\Pi \rvert_{t = 0}\equiv 0\).
The scalar equation~\eqref{eq: scalar equation sGB} then becomes
\begin{multline}\label{eq: scalar equation normal formulation}
    \partial_{i}  \left(\gamma^{ij}\partial_j \Psi\right) + \gamma^{ij}\partial_j \Psi
    \left(\partial_i \ln\alpha+\Gamma^{k}_{ki}\right) = - \ell^2 f'(\Psi) \mathcal{G},
\end{multline}
where \(\Gamma^{k}_{ij}\) is the 3-dimensional spatial Christoffel symbol with respect to \(\gamma_{ij}\).
Equation~\eqref{eq: scalar equation normal formulation} is both elliptic and regular everywhere.
In Ref.~\cite{Kovacs:2021lgk}, the inner boundary $S_\mathrm{in}$ is placed on the apparent horizon of the BH and is supplemented with boundary conditions at both inner and outer boundaries,
\begin{align} \label{eq: Normal formulation boundary conditions I}
    \left.\hat{s}^i \partial_i \Psi\right|_{S_{\mathrm{in}}} & = 0,              \\
    \lim _{r \rightarrow \infty} \Psi                        & = \Psi_{\infty} ,
\end{align}
where \(\hat{s}^i\) is the unit outward normal vector to the BH horizon(s).
For computational domains extending inside the apparent horizon, we instead impose  a constant Dirichlet boundary condition (i.e., $\left.\Psi\right|_{S_\mathrm{in}}=\mathrm{const.}$), chosen such that $\hat{s}^i \partial_i \Psi = 0$ on each apparent horizon.
On a finite spatial domain, and assuming an asymptotic decay of the scalar of the form of Eq.~\eqref{eq: Psi asymptotic dacay}, we replace the outer boundary condition with [c.f.~Eq.~\eqref{eq: decay boundary condition}] a Robin type boundary condition
\begin{align}\label{eq: outer bc}
    \left( r \hat{s}^{i}\partial_{i} \Psi + \Psi - \Psi_{\infty} \right)\Big|_{S_\text{out}} = 0 ~,
\end{align}
where \(\hat{s}^i\) is now the unit outward normal vector to the outer spherical boundary \(S_\text{out}\). We set \(\Psi_\infty \equiv 0\). We shall refer to this as the ``$\partial_n=0$'' formulation.


\subsubsection{Caveats of the normal formulation}

While the $\partial_n=0$ formulation provides a readily solvable elliptic system, the most common use case for the XCTS formulation is the calculation of quasiequilibrium initial conditions. Unfortunately, the normal formulation will not generically lead to stationary spacetimes.
Consider, for example, the case of a Schwarzschild BH in Kerr-Schild coordinates [Eq.~\eqref{eq: KerrSchild metric spherical symmetry}]. The timelike Killing vector \(\xi\) of the spacetime is
\begin{align}\label{eq: time}
    \xi^{a} = t^{a} = \alpha n^{a} + \beta^{a}.
\end{align}
Assuming that the momentum $\Pi$ is initially zero, the initial time derivative of the scalar field $\Psi$ is
\begin{align} \label{eq: violation of stationarity}
    t^{c} \partial_{c}\Psi = \alpha n^{c} \partial_{c} \Psi + \beta^{i} \partial_{i} \Psi = \beta^{i } \partial_{i}\Psi.
\end{align}
Therefore, whenever $\beta^i\neq0$ and $\partial_i \Psi \neq0$, \(\mathcal{L}_{\xi} \Psi\) will not vanish. For our example, \(\mathcal{L}_{\xi} \Psi = \beta^{r}\partial_r \Psi = 2M/(r + 2M)\partial_r \Psi \neq 0\)
and the scalar hair obtained will not be stationary.
Indeed, solving the spherically symmetric version of Eq.~\eqref{eq: scalar equation normal formulation} in our 1D code, we find a profile different from the $\partial_t=0$ solution constructed in Sec.~\ref{sec: theory solving 1d problem}.  This profile is also shown in Fig.~\ref{fig: 1d profiles}.

Finally, we note that the inner boundary condition [Eq.~\eqref{eq: Normal formulation boundary conditions I}] is inconsistent with stationarity.
Indeed, if \(\Psi\) is regular at the horizon, then it can be expanded as a series about \(r_h\) of the form
\begin{align}
    \Psi (r) = \sum_{n = 0}^{\infty} \Psi_{(i)} \left(r - r_h\right)^n.
\end{align}
Solving Eq.~\eqref{eq: scalar equation in KerrSchild} order-by-order perturbatively in \(\Delta r = r-r_h\), we obtain that
\begin{align}
    \partial_r \Psi \lvert_{r_h} = \Psi_{(1)} = -\frac{3}{8M^3}\ell^{2}\left(\eta\Psi_{(0)}+\zeta\Psi_{(0)}^{3}\right),
\end{align}
which is nonzero in general, contradicting Eq.~\eqref{eq: Normal formulation boundary conditions I}.


\subsection{\boldmath 3D approximate Killing formulation ``$\partial_t=0$''} \label{sec: approximate Killing formulation}

Motivated by the existence of a symmetry along a Killing vector, we present a new procedure for extending the XCTS formulation to sGB gravity, which we refer to as the ``$\partial_t=0$'' formulation.
The main assumption will now be that the ``momentum'' with respect to the (approximate) Killing vector \(\xi^{a}\), given by
\begin{align}
    P \equiv  \mathcal{L}_\xi \Psi,
\end{align}
vanishes on the initial slice.

From the previous discussion, for a stationary GR black hole in coordinates adapted to the symmetry, as well as for solutions of the XCTS equations, the Killing vector corresponds to \(\xi = \partial_t\).
By imposing \(P  = \partial_t \Psi \equiv 0 \), Eq.~\eqref{eq: scalar equation sGB} becomes
\begin{equation} \label{eq: scalar equation svbp}
    \partial_{i}  \left(\mathbb{M}^{ij}\partial_j \Psi\right)
    +     \left(\partial_i \ln\alpha+\Gamma^{k}_{ki}\right)\mathbb{M}^{ij}\partial_j \Psi
    =
    - \ell^2 f'(\Psi) \mathcal{G},
\end{equation}
where
\begin{equation} \label{eq: principal part of sbvp equation}
    \mathbb{M}^{ij} \equiv \gamma^{ij} - \alpha^{-2}
    \beta^{i} \beta^{j}.
\end{equation}
Equation~\eqref{eq: scalar equation svbp} is the 3D generalization of
Eq.~\eqref{eq: scalar equation in KerrSchild}.
In the spirit of quasiequilibrium, we have also set
\(\partial_t \alpha = 0\) and \(\partial_t \beta^{i} = 0\)
in the derivation of Eq.~\eqref{eq: scalar equation svbp}.  We note that these simplifications could be relaxed and their values can be set according to a desired gauge choice.

The principal part of Eq.~\eqref{eq: scalar equation svbp} is $\mathbb{M}^{ij}\partial_i\partial_j\Psi$.
The singularity at $r_h$ in the 1D formulation [Eq.~\eqref{eq: scalar equation in KerrSchild}] now
corresponds to the situation where
\begin{align}
    \left.\det \mathbb{M} \right\lvert_{S_h} = 0,
\end{align}
i.e., when (at least) one of the eigenvalues of \(\mathbb{M}^{ij}\) vanishes at the apparent horizon \(S_h\).
$\mathbb{M}$ is singular on the BH horizon in general.  For a stationary BH in time-independent coordinates, the time vector on the horizon must be parallel to the horizon generators as argued in Ref.~\cite{Cook:2004kt}, which implies that on the horizon $\beta^i \hat{s}_i = \alpha$, where $\hat{s}_i$ is the outward-pointing spatial unit normal to the horizon. Using this equality, it follows that $\hat{s}_i \mathbb{M}^{ij} \hat{s}_j=0$.

As for the spherically symmetric example above, our approach
will be to rely on the inherent smoothness of spectral expansions to
single out solutions of Eq.~\eqref{eq: scalar equation svbp} that
smoothly pass through the horizon.  Regularity at the horizon
reduces the number of possible solutions, and so we will not impose
a boundary condition at the excision surface in the interior of the
horizon.  We note that Lau et al.~\cite{Lau:2008fb,Lau:2011we} encountered the same principal part as Eq.~\eqref{eq: scalar equation svbp} in the context of IMEX evolutions on curved backgrounds. Reference~\cite{Lau:2008fb} in particular contains an analysis of the singular boundary value problem. We impose the boundary condition (\ref{eq: outer bc}) at the outer boundary, where again we set \(\Psi_\infty \equiv 0\). Note that, in spherical symmetry, using $\gamma^{ij}$, $\alpha$, and $\beta^{i}$ corresponding to that of the Kerr-Schild metric, this formulation reduces to Eq.~\eqref{eq: scalar equation in KerrSchild}.

\begin{figure}
    \includegraphics[width=\linewidth]{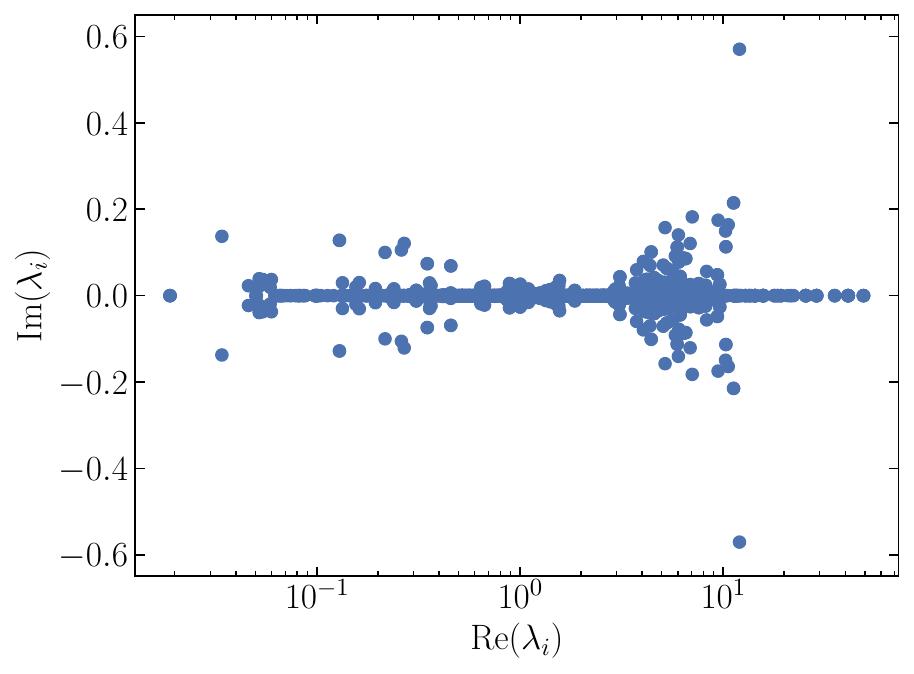}
    \caption{
        \textit{Behavior of elliptic problem at horizon}
        Eigenvalues of the linearized discretized operator of Eq.~\eqref{eq: scalar equation svbp} for a discontinuous Galerkin element that crosses the horizon. The eigenvalues all have a positive real part and span $\sim3$ orders of magnitude, showing that the matrix is invertible and moderately well-conditioned.}
    \label{fig: eigenvalues 3d sbvp}
\end{figure}


\subsection{3D numerical implementation}
\label{sec:3Dimplementation}

To solve the nonlinear Eqs.~\eqref{eq: scalar equation normal formulation} and \eqref{eq: scalar equation svbp} in 3 dimensions, we employ the spectral elliptic solver~\cite{Vu:2021coj} of the open-source \texttt{SpECTRE} code~\cite{deppe_2024_11494680}. \texttt{SpECTRE} employs a discontinuous Galerkin discretization scheme, where the domain is decomposed into elements, each a topological \(d\)-dimensional cube. These elements do not overlap but share boundaries. Boundary conditions on each element (both external boundary conditions, as well interelement boundaries)  are encoded through fluxes. We refer the reader to Refs.~\cite{Fischer:2021voj,Vu:2021coj} for more details about the mathematical formulation and numerical implementation.

For our present study of Eqs.~\eqref{eq: scalar equation normal formulation} and \eqref{eq: scalar equation svbp} in the decoupling limit, $\mathcal{G}$ is known and nonlinearities enter only through $f'(\Psi)$.
Since the full linearization of these equations in \(\Psi\) is straightforward, we solve them by utilizing the Newton-Raphson algorithm within \texttt{SpECTRE}.

In general, in the fully-coupled system [\(H_{ab} \neq 0\) in Eq.~\eqref{eq: tensor equation}], additional terms enter the original XCTS equations and the full linearization strategy described above becomes impractical: first, because one would need to linearize in both the scalar \emph{and} metric variables; and second, because such nonlinearities are very specific to the concrete theory. Indeed, in the case of sGB, these arise from the intricate structure of both \(\mathcal{G}\) and \(H_{ab}\), which depend on (up to second-order derivatives of) the scalar and metric variables.
To avoid a large implementation burden, and explore possible strategies for future work,
we also implement a straightforward over-relaxation scheme, which can easily be extended to other theories. Note that a similar relaxation scheme was recently employed in Ref.~\cite{Brady:2023dgu}.

Our relaxation scheme constructs increasingly accurate approximants $\Psi^{(K)}$, $K=1, 2, \ldots$, to the solution, where in each iteration $K$ the nonlinearity is calculated from earlier iterations.  Specifically, for the scalar equation~\eqref{eq: scalar equation svbp}, we solve
\begin{multline}\label{eq: damping eq}
    \partial_{i}  \left(\mathbb{M}^{ij}\partial_j \Psi^{(K)}\right)
    + \mathbb{M}^{ij}\partial_j \Psi^{(K)} \left(\partial_i \ln\alpha+\Gamma^{k}_{ki}\right) \\
    = - \ell^2 f'(U^{(K)}) \mathcal{G}.
\end{multline}
with
\begin{align}\label{eq: damping field}
    U^{(K)} & = \varepsilon \Psi^{(K-1)} + (1-\varepsilon)U^{(K-1)}, & K\ge 1, \nonumber
    \\ U^{(0)} &= \Psi^{(0)},& K=0.
\end{align}
Here \(\varepsilon \in [0, 1]\) is a damping parameter, \(\Psi^{(0)}\) is the initial guess, and an analogous expression holds for Eq.~\eqref{eq: scalar equation normal formulation}.
Upon discretization, at each iteration \(K\), a linear problem of the form
\begin{align}
    \mathbb{A} \boldsymbol{y} = \boldsymbol{b}
\end{align}
is solved for \(\boldsymbol{y} = \{\Psi^{(K)}(x_i)\}\), with \(x_i\) being the nodal points of the spectral basis consisting of tensor products of Legendre polynomials.
Here, \(\boldsymbol{b}\) is a fixed source term which only depends on quantities of the previous iteration \(K\!-\!1\).
Boundary conditions are imposed through the discontinuous Galerkin fluxes, ensuring that the matrix \(\mathbb{A}\) is invertible.
Since the Legendre polynomials are finite and regular within each element, regularity across the horizon is guaranteed so long as the horizon does not coincide with element boundaries.
The scheme~\eqref{eq: damping eq} is iterated until the residual of Eq.~\eqref{eq: scalar equation svbp} or~\eqref{eq: scalar equation normal formulation} is sufficiently small. For all solves presented here, we use a tolerance of $10^{-10}$.

\begin{figure}
    \includegraphics[width=0.96\linewidth]{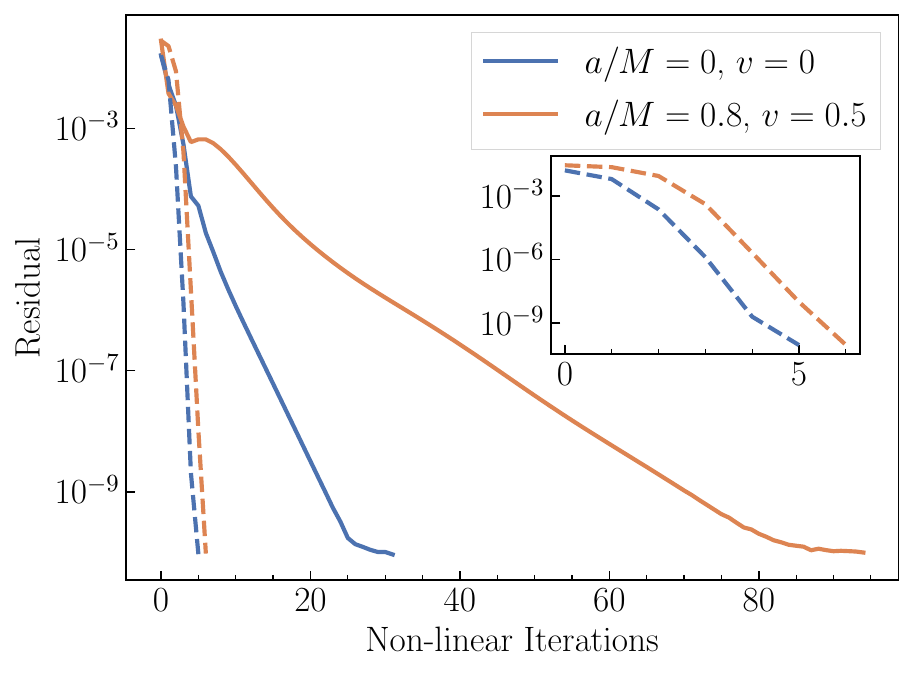}
    \caption{\label{fig: convergence}
    \textit{Performance of iterative numerical scheme.}
    Residual at each nonlinear iteration for a solve on a nonrotating BH background, as well as a boosted, rotating BH background. In both cases we set $\ell^2\eta/M^{2}=6$, $\ell^2\zeta/M^{2}=-60$,
    and in the latter the BH has a speed of \(v = 0.5\) in the \(x\)-direction, and dimensionless spin of $a/M = 0.8$ in the \(z\)-direction. The dashed lines indicate the same solves, but using the full linearization and a Newton-Raphson algorithm, for comparison.
    }
\end{figure}

To further demonstrate the well-posed nature of the elliptic equation~\eqref{eq: scalar equation svbp}, in Fig.~\ref{fig: eigenvalues 3d sbvp}, we plot the eigenvalues of the submatrix of $\mathbb{A}$ in an element crossing the horizon of the BH. All eigenvalues are nonzero, indicating the matrix is invertible.  Furthermore, all eigenvalues have positive real parts, indicating this matrix should be amenable to standard iterative linear solvers.
The real parts of the eigenvalues span $\sim 3$ orders of magnitude, indicating that the matrix is moderately well-conditioned, and numerically we are able to invert the linear system without problems.


\section{Results: single black holes} \label{sec: single black holes results}

\subsection{3D code in spherical symmetry}\label{sec: single BH 3D}

We will now apply the formalism and code developed above to
spacetimes with a single black hole.  We start with spherical symmetry, where we solve the scalar equation for coupling
constants $\ell^2\eta=6M^2$ and $\zeta=-10\eta$ within the
``$\partial_t=0$'' formulation [Eq.~\eqref{eq: scalar equation svbp}], in both the 1D and 3D code (the
1D result is shown in Fig.~\ref{fig: 1d profiles}).
Figure~\ref{fig: convergence} showcases the convergence of our numerical implementation of the 3D initial data. The figure shows the convergence with
iteration number of the full Newton-Raphson scheme and the
relaxation scheme~\eqref{eq: damping eq}.  While the full
Newton-Raphson scheme converges more quickly, the relaxation
scheme also works reliably and reasonably efficiently.

\begin{figure}
    \includegraphics[width=0.96\linewidth]{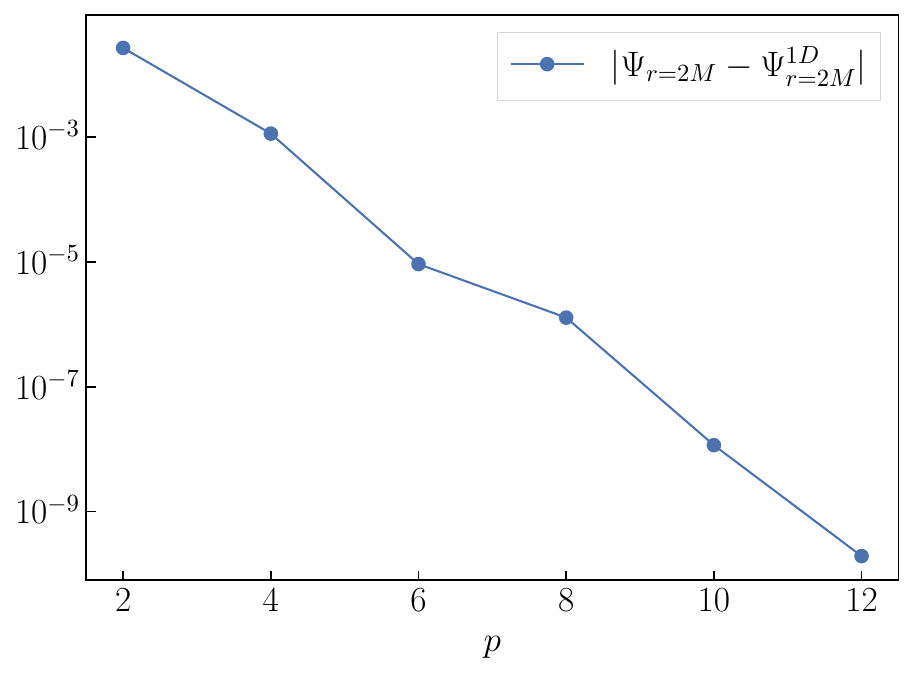}
    \caption{\textit{Comparison 1D vs.~3D code for a scalarized Schwarzschild BH.}  Plotted is the absolute difference between the value of the scalar field $\Psi$ at the horizon between the 1D code and 3D code, for varying polynomial order $p$ in the 3D code. The solution $\Psi$ is plotted in Fig.~\ref{fig: 1d profiles}.
    }
    \label{fig: 1d convergence}
\end{figure}

Turning to the accuracy of these spherically symmetric numerical solutions, we compare our 3D \texttt{SpECTRE} implementation with the 1D \texttt{Python} code presented in Sec.~\ref{sec: theory solving 1d problem}.
We solve the scalar  equation and compute the value of the scalar field at the horizon.
Figure~\ref{fig: 1d convergence} shows the difference between the two codes as a function of the resolution in the 3D code. We find that the 3D code converges to the same answer exponentially, and achieves an accuracy of better than $10^{-9}$.

\subsection{3D code without symmetries}

\begin{figure*}[]
    \includegraphics[width=0.45\linewidth,trim=0 5 0 5]{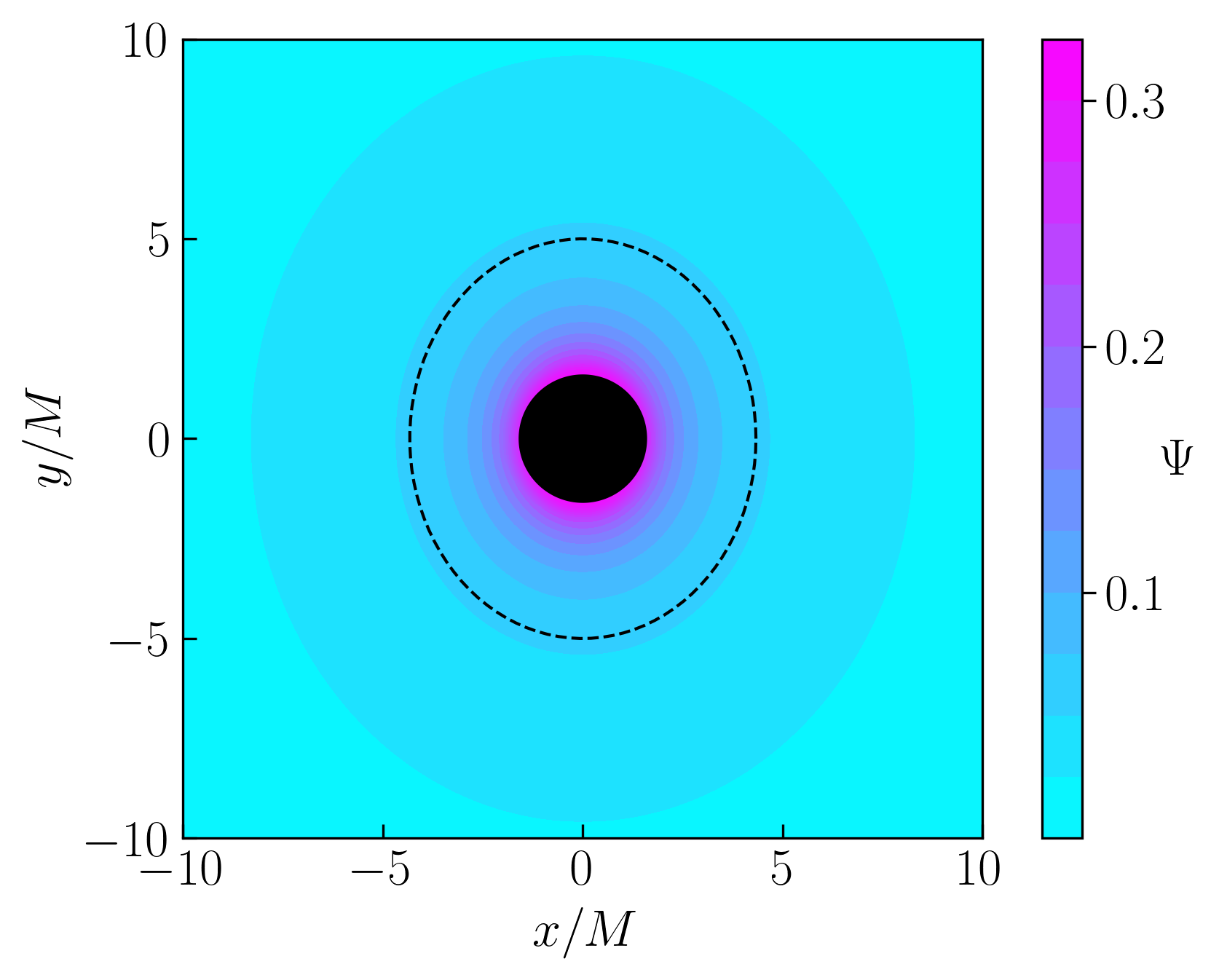}
    $\quad$
    \includegraphics[width=0.47\linewidth]{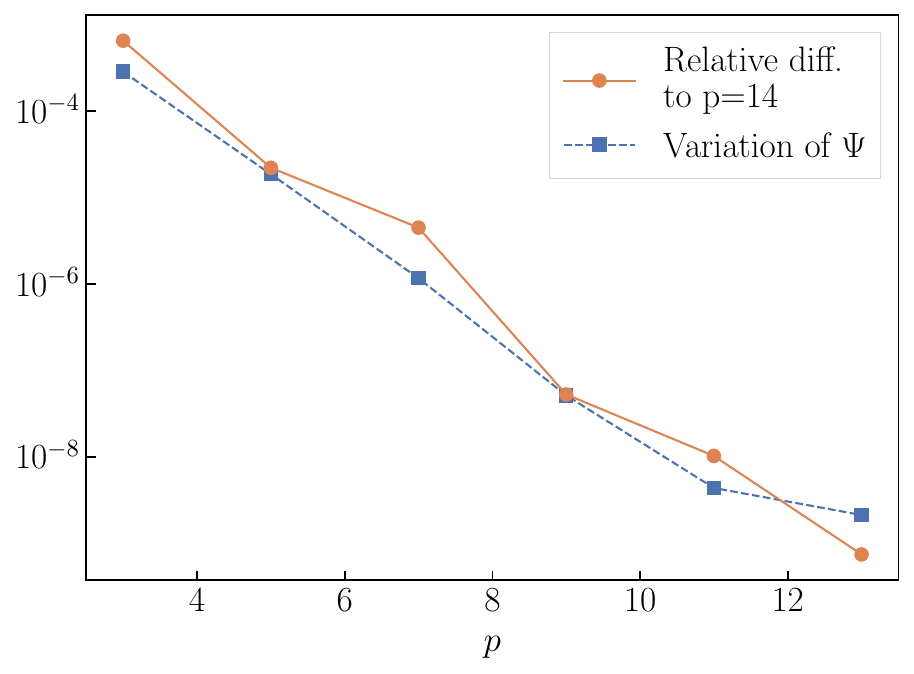}
    \caption{\textit{Scalar field for a boosted, rotating BH.} The background metric corresponds to a BH with spin $a/M=0.8$ (in the $z$-direction), boosted with \(v = 0.5\) in the \(x\)-direction, for coupling parameters
    $\ell^2\eta/M^{2}=6$ and $\zeta=-10\eta$.
    \textbf{Left}: Contour plot of the scalar field $\Psi$ in the \(xy\)-plane. The black disk at the center represents the inner excision region, while the dashed ellipse is used for the convergence plot on the right. \textbf{Right}: convergence test. The orange line shows the \(L_2\)-norm of the difference between each resolution and highest resolution ($p=14$) on the dashed ellipse. The blue line
    demonstrates that the solution is constant on the dashed ellipse by
    plotting the \(L_2\)-norm of the difference of the scalar field around the ellipse compared to the average value on it.}
    \label{fig: boost}
\end{figure*}

We now consider a genuinely nonsymmetric 3-dimensional configuration:  a black hole with spin \(a/M = 0.8\) along the \(z\)-axis, boosted to velocity $v=0.5$ in the direction of the \(x\)-axis. The background spacetime is given in Cartesian Kerr-Schild coordinates \(\boldsymbol{x} = (x, y, z)\) as
\begin{align}\label{eq:KerrSchild}
    g_{ab} = \eta_{ab} + 2\mathcal{H} l_a l_b.
\end{align}
Here \(\eta_{ab} = \mathrm{diag}(-1, 1, 1, 1)\) is the Minkowski metric, and
the scalar function \(\mathcal{H}\) and one-form \(l_a\) (which satisfies \(l^c \partial_c l_a = l^c \nabla_c l_a = 0\)) are given by
\begin{align}\label{eq:KerrSchild2}
    \mathcal{H} & \equiv  \dfrac{M \rho^3}{\rho^4 +a^2 z^2}\,,                                                          \\
    l_a         & \equiv \left(1, \dfrac{\rho x+ay}{\rho^2+a^2}, \dfrac{\rho y-ax}{\rho^2+a^2}, \dfrac{z}{\rho}\right),
\end{align}
with \(\rho\) implicitly defined through
\(\rho^2(x^2+y^2) + (\rho^2+a^2)z^2 = \rho^2(\rho^2+a^2)\),
and $M$ and $a$ being the BH mass and spin parameter, respectively.

The background Eq.~\eqref{eq:KerrSchild} is boosted by applying the appropriate Lorentz boost to the coordinates \(x^a\) and the null vector \(l_a\). We apply a Galilean transformation to the shift, i.e.~\(\beta^{i} \to \beta^{i} + v^{i}\), where \(v^{i}\) is the boost velocity of the BH, to obtain stationary coordinates.

We now solve Eq.~\eqref{eq: scalar equation svbp} on this background with the same coupling constants as above,
$\ell^2\eta=6M^2$ and $\zeta=-10\eta$.
Our numerical scheme successfully solves the singular boundary value problem even in this more complex configuration, although Fig.~\ref{fig: convergence} shows an increase in the number of relaxation/nonlinear iterations.

The left panel of Fig.~\ref{fig: boost} shows the spatial dependence of the calculated scalar field $\Psi$ in the \(xy\)-plane.
The coupling parameters are the same as above, while the BH has
dimensionless spin $a/M=0.8$ and a boost velocity of \(v = 0.5\) in the \(x\)-direction.
The scalar field is largest near the black hole and falls off at large distance.
The boost manifests itself as a length contraction along the direction of the velocity, which can be seen by the shape of the contour lines.  As a guide to the eye, a dashed ellipse in the left panel of Fig.~\ref{fig: boost} is plotted with the correct Lorentz contraction for $v=0.5$.

The right panel of Fig.~\ref{fig: boost} presents two different convergence tests for the scalar field values on the dashed ellipse of the left panel. First, we compare the values along the ellipse at polynomial resolution $p$ to those obtained in our highest resolution solution with $p_{\rm max}=14$.  We plot this difference vs $p$ and find exponential convergence.
Second, because the boost direction and the spin direction are orthogonal,
we expect the scalar field to be constant on the dashed ellipse in the left panel.  We test this expectation by computing at each resolution $p$ the variance of $\Psi$ along the ellipse and plot it vs $p$ in the right panel.
We find that this variance decays exponentially to zero with increasing resolution $p$.

\begin{figure*}[]
    \includegraphics[width=0.45\linewidth,trim=0 10 0 10]{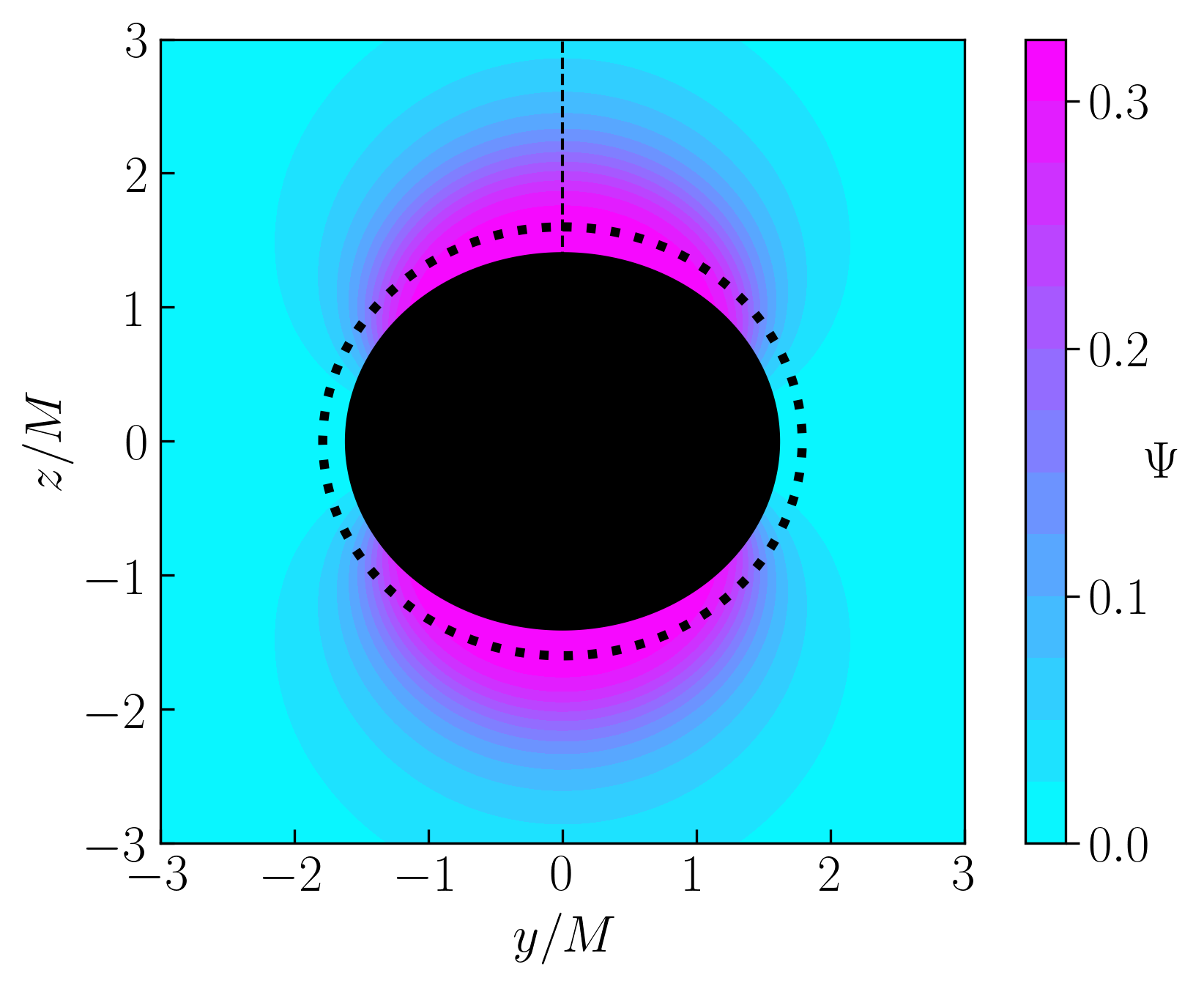}
    $\quad$
    \includegraphics[width=0.47\linewidth]{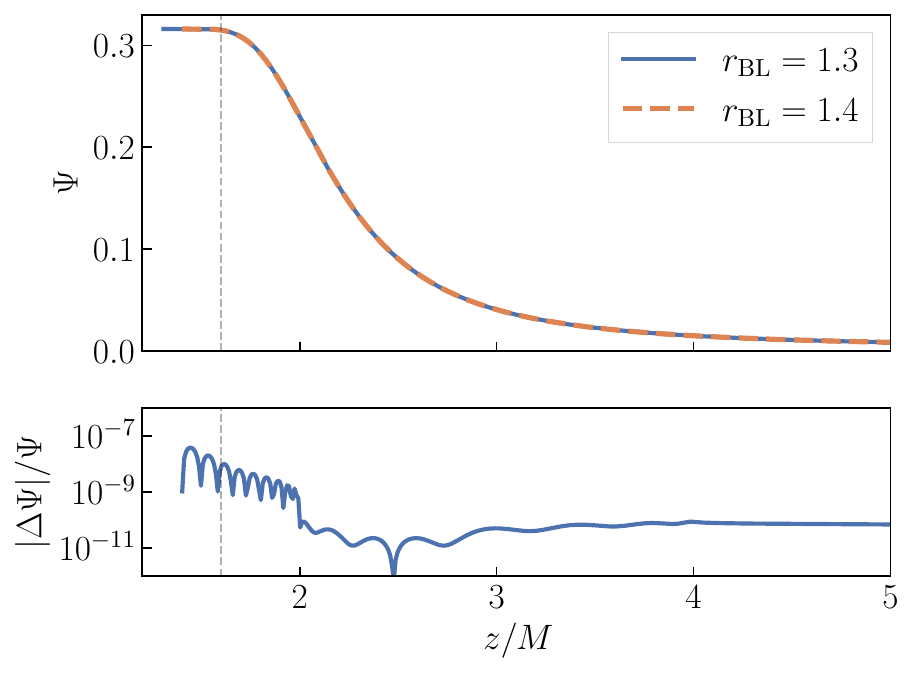}
    \caption{\textit{Spin-induced scalar profile.} The background metric corresponds to a BH with spin $a/M=0.8$, with coupling parameters $\ell^2\eta/M^{2}=-40$ and $\zeta=-10\eta$. \textbf{Left}: contour plot of the scalar field $\Psi$ in the \(yz\)-plane. The black disk at the center represents the inner excision region, while the dotted line indicates the BH horizon. The dashed line is used for the test in the right panel. \textbf{Right}: test of the impact of inner excision surface radius on scalar profile. Plotted are two scalar field solves, where the Boyer-Lindquist radius $r_{\rm BL}$ of the inner exicision surface is varied. The bottom panel shows the relative difference between the two profiles.
    }
    \label{fig: spin_induced}
\end{figure*}

As a final test for single BH spacetimes, we demonstrate that our code is
also capable of obtaining scalar profiles resulting from \emph{spin-induced} scalarization~\cite{Dima:2020yac, Doneva:2020nbb}.
For black holes rotating rapidly enough, the Gauss-Bonnet scalar $\mathcal{G}$ can change sign near the poles of the BH.
This allows us to choose a negative coupling \(\eta<0\) to source the scalar field in such regions--note that, for \(\eta <0\), a nonrotating BH would \emph{not} acquire scalar hair as \(m_{\Psi, \mathrm{eff}} > 0\) everywhere. The left panel of Fig.~\ref{fig: spin_induced} shows one such profile obtained with the ``$\partial_t=0$'' formulation, with $\ell^2\eta/M^{2}=-40$ and $\zeta=-10\eta$, and $a/M=0.8$. We use this system to perform a further test of the uniqueness of solutions obtained using this formulation: we perform two scalar field solves for which we vary the Boyer-Lindquist radius $r_{\rm BL}$ of the inner excision surface inside the BH horizon. The right panel of Fig.~\ref{fig: spin_induced} compares these solutions along the \(z\)-axis, with the lower panel showing the relative difference between two. This deviation is within the truncation error of the numerical solutions, further supporting our claim that the solution is fully determined without specifying a boundary condition at the inner excision surface.

\subsection{Evolution of scalar field initial data}

Finally, we evolve the 3D initial data sets in the decoupling limit.
We evolve single BH initial data within \texttt{SpECTRE} with the code described in Ref.~\cite{Lara:2024rwa}.  For initial data corresponding to the approximate Killing formulation (Sec.~\ref{sec: approximate Killing formulation}),
we complete the initial data set by computing the momentum \(\Pi\) [Eq.~\eqref{eq: momentum Pi definition}] as
\begin{align}
    \Pi \lvert_{t = 0} = \alpha^{-1}\beta^i \partial_i \Psi,
\end{align}
while for the ``$\partial_n=0$'' formulation we set $\Pi\lvert_{t = 0}=0$, consistent with the assumptions of this formulation.
The evolution equations are discretized with a discontinuous
Galerkin scheme employing a numerical upwind flux~\cite{Kidder:2016hev}.  Time evolution is
carried out by means of a fourth-order Adams-Bashforth time-stepper
with local adaptive time-stepping~\cite{throwe2020highorder}, and we
apply a weak exponential filter on all evolved fields at each time
step~\cite{Hesthaven}.
For the evolution of the metric variables, we use a generalized harmonic system~\cite{Lindblom:2005qh} with analytic gauge-source function \(H^{c} = {^{(4)}\Gamma^{c}}\), where \({^{(4)}\Gamma^{c}} = g^{ab} {^{(4)}\Gamma^{c}_{ab}}\) is a contraction of the 4-dimensional Christoffel symbol computed from Eq.~\eqref{eq:KerrSchild}.
The spatial domain consists of a series of concentric spherical
shells with outer boundary located at \(R / M = 500\). A region inside
the BH is excised and the inner boundary conforms to the shape of the
apparent horizon.

\begin{figure}
    \includegraphics[width=0.96\linewidth]{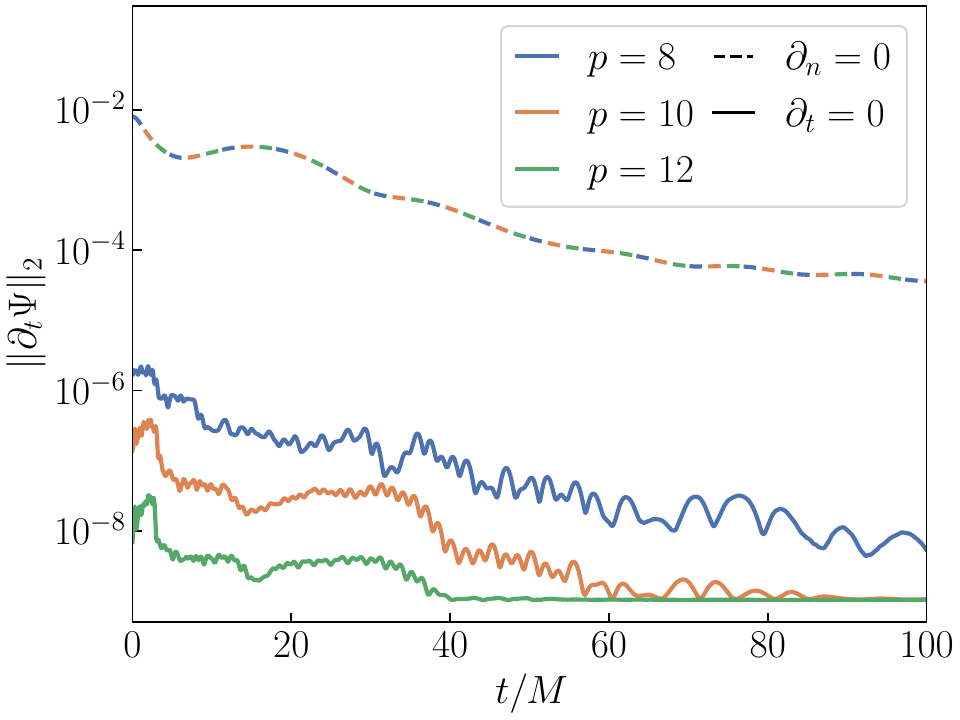}
    \caption{\textit{Evolution of initial data for a nonrotating BH.} \(L_2\)-norm over the entire domain of the time derivative of $\Psi$ for initial data generated via the ``$\partial_n=0$'' formulation (dashed curves) and the ``$\partial_t=0$'' formulation (solid curves) for varying grid resolution \(p\). Physical system is the same as that in Fig. \ref{fig: 1d profiles}.
    }
    \label{fig: rhs}
\end{figure}

Figure~\ref{fig: rhs} shows the time derivative of the scalar profile
for early parts of the evolution.
With increasing
initial data resolution (larger \(p\)), the initial dynamics for the
``$\partial_t=0$'' formulation decreases,
whereas for the
``$\partial_n=0$'' case it remains large. This behavior confirms our earlier
findings: only the $\partial_t=0$ formulation in Eq.~\eqref{eq: Normal
    formulation boundary conditions I} yields time-independent scalar
field configurations.

\section{Binary black hole hair} \label{sec: binary black holes}

In this section, we present quasistationary hair configurations for black hole binaries using the ``$\partial_t=0$'' formulation described in Sec.~\ref{sec: approximate Killing formulation}.

\subsection{Background spacetime}

For binary BHs,
we obtain numerical background solutions by solving the XCTS system of equations in \texttt{SpECTRE} for a binary black hole system.
We choose the conformal metric \(\bar{\gamma}_{ij}\) and extrinsic curvature \(K_{ij}\) as superposed Kerr-Schild data~\cite{Matzner:1998pt,Marronetti:2000yk,Lovelace:2008tw} and solve the XCTS equations with the code presented in Ref.~\cite{Vu:2021coj}.
The numerical solution is then imported into our scalar field solver.

To avoid rank-4 tensors, the Gauss-Bonnet invariant \(\mathcal{G}\) is computed (in vacuum) from the background metric in terms of the electric \(E_{ij}\) and magnetic \(B_{ij}\) parts of the Weyl scalar as
\begin{align}\label{eq:G_from_EB}
    \mathcal{G} = 8 \left(E_{ij} E^{ij} - B_{ij} B^{ij}\right).
\end{align}
We refer the reader to Ref.~\cite{Okounkova:2017yby} for the definitions of these quantities.

\subsection{Light cylinder}
\label{sec: binary black hole hair light cylinder issue}

For a BH binary, with orbital frequency \(\boldsymbol{\Omega} = \Omega \, \hat{\boldsymbol{z}}\), we can decompose the shift into
\begin{align}\label{eq:shift-decomp}
    \boldsymbol{\beta} = \boldsymbol{\Omega}\! \times\! \boldsymbol{r}+\boldsymbol{\beta}_\text{(exc)},
\end{align}
where the first term describes the corotation of the coordinates with the binary and \(\boldsymbol{\beta}_\text{(exc)}\) is the  \textit{shift excess} \cite{Vu:2021coj} solved for in the XCTS equations.
Because $\boldsymbol{\Omega} \times \boldsymbol{r}$ grows without bound for large $r$, and because $\boldsymbol{\beta}_{\rm (exc)}$ is finite, the shift can achieve magnitudes \(\lvert \boldsymbol{\beta}\rvert \gtrsim 1 \).
As the shift appears in the principal part of Eq.~\eqref{eq: scalar equation svbp}, the superluminal coordinate velocity
leads to a change in character of Eq.~\eqref{eq: scalar equation svbp} from elliptic to hyperbolic.
To illustrate this more clearly note that \(\gamma^{ij}\) and \(\alpha\) asymptote to the Kronecker delta \(\delta^{ij}\) and 1, respectively. Writing the shift as \(\boldsymbol{\beta} = (-\Omega y, \Omega x, 0)\), the three eigenvalues of the matrix \(\mathbb{M}^{ij}\) [Eq.~\eqref{eq: principal part of sbvp equation}] are
\begin{align}
    \lambda_{1, 2} = 1 \quad \text{and} \quad \lambda_{3} = 1 - \Omega^2 (x^2 + y^2) ~.
\end{align}
For cylindrical radius \(\varrho \equiv \sqrt{x^2 + y^2} < 1 / \lvert \Omega \rvert \), all eigenvalues are positive and Eq.~\eqref{eq: scalar equation svbp} is elliptic. Instead, for \(\varrho \geq 1 / \lvert \Omega \rvert\), Eq.~\eqref{eq: scalar equation svbp} is either parabolic or hyperbolic. The boundary
\begin{align}
    \varrho_\text{LC} \equiv \dfrac{1}{\lvert \Omega \rvert}
\end{align}
is called the \textit{light cylinder}--see, e.g., Ref.~\cite{Klein:2004be}.

These considerations are indeed relevant in practice for solving for binary BHs: numerically, we find that if the outer boundary of the domain is within the light cylinder, the numerical solver converges, whereas, if it is beyond the light cylinder, the solver does not converge.
We conclude that for Eq.~\eqref{eq: scalar equation svbp} with nonzero orbital velocity on a large domain our numerical methods are no longer guaranteed to be effective.

To restore ellipticity of Eq.~\eqref{eq: scalar equation svbp}, we introduce a spherical \textit{roll-off} function on the terms involving the shift. That is, we replace Eq.~\eqref{eq: scalar equation svbp} by
\begin{multline}\label{eq: sbvp with rolloff}
    \partial_{i}  \left(\left[\gamma^{ij}-F(r)\alpha^{-2}\beta^i\beta^j\right]\partial_j \Psi\right)  \\
    + \left[\gamma^{ij}-F(r)\alpha^{-2}\beta^i\beta^j\right]\partial_j \Psi \left(\partial_i \ln\alpha+\Gamma^{k}_{ki}\right) \\
    = - \ell^2 f'(\Psi) \mathcal{G}.
\end{multline}
The roll-off function
\begin{align}
    F(r) \equiv \dfrac{1}{2} \left\{1 - \tanh \left[\mu (r - r_\text{roll-off})\right]\right\}
\end{align}
depends on shape parameters $\mu$ and $r_\text{roll-off}$, which adjust the width and location of the roll-off, respectively. With a roll-off inside the light cylinder, our numerical solver converges without problems.

\begin{figure}
    \includegraphics[width = \linewidth]{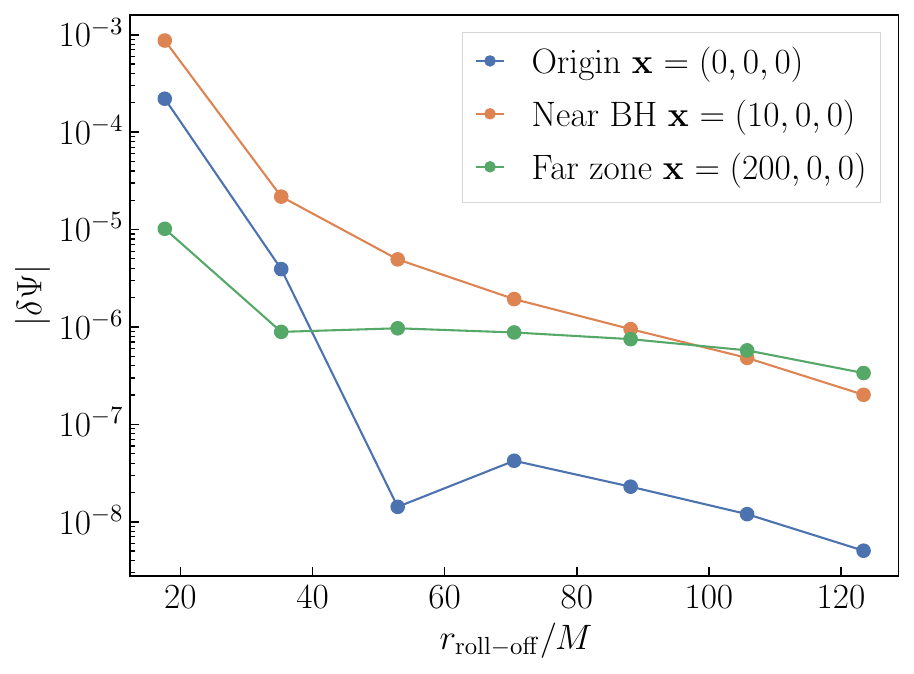}
    \caption{\textit{Impact of roll-off.}
    We consider $\delta\Psi\equiv \Psi(\mathbf{x})_{r_{\rm roll-off}} -  \Psi(\mathbf{x})_{r_{\rm roll-off}\simeq141M}$ for binary BH solutions with different $r_{\rm roll-off}$. Different lines correspond to comparison at different regions of the computational domain. The binary considered is the same as that in the right panel of Fig.~\ref{fig: binary}, with the black holes placed at $(\pm 8,0,0)$.}
    \label{fig: rolloff}
\end{figure}

Because the rolled-off shift terms are primarily in angular directions [cf.~Eq.~\eqref{eq:shift-decomp}], we expect that the inclusion of \(F(r)\) will lead to some loss of angular structure beyond the roll-off radius.
Since the rolled-off region is placed relatively far from the binary, we expect a marginal impact from this on the  dynamics.
To quantify the impact of the roll-off, we solve Eq.~\eqref{eq: sbvp with rolloff} for different values of $r_\text{roll-off}$. Figure~\ref{fig: rolloff} shows the variation of the scalar field at representative points near and far from the BHs: the origin (where $\Psi\simeq 0.0536$), a point very near to a BH horizon (where $\Psi\simeq 0.1097$) and a point in the far zone (where $\Psi\simeq 0.0026$). The solutions are obtained  with a numerical accuracy of $\sim 10^{-8}$, corresponding to $p=7$ of the convergence test we discuss next. Even in the far-field, where $F(r)=0$, the fractional change in $\Psi$ is less than $10^{-3}$; near the black holes, the fractional change is below $10^{-5}$. Therefore, we believe that the inclusion of the roll-off factor should have a very limited effect on the dynamics.

\subsection{Scalar hair around binary black holes}

Finally, in Fig.~\ref{fig: binary}, we present the scalar profile induced by a binary black hole system. The black holes are both nonspinning, with mass $M$,
and are in an approximately quasicircular configuration with $\Omega\simeq0.0082/M$, placing the light cylinder at $\rho_{\rm LC}\simeq122M$. The coupling constants were chosen as \(\ell^2\eta/M^{2}=3.34\) and \(\ell^2\zeta/M^{2}=-31.1\).

Both solutions displayed in Fig.~\ref{fig: binary} are solutions to the \emph{same} boundary-value problem [Eq.~\eqref{eq: scalar equation svbp} with boundary condition~\eqref{eq: outer bc}] on an identical background geometry.  This illustrates the nonuniqueness of solutions to this nonlinear problem; in fact, two more solutions can be obtained by $\Psi\to -\Psi$.  Which solution is obtained can be controlled by the choice of initial guess $\Psi^{(0)}$ for the relaxation scheme described in Sec.~\ref{sec:3Dimplementation}.
In order to obtain the solution with like charges, we chose our initial guess as a superposition of two $A/r$ profiles centered on each BH. To obtain the solution with opposite sign charges, we flip the sign of one of the $A/r$ terms in the initial guess. The scheme is not sensitive to the precise coefficients $A$ in the $1/r$ profiles.

\begin{figure}
    \includegraphics[width=0.9\linewidth,clip=true,trim=0 7 0 7]{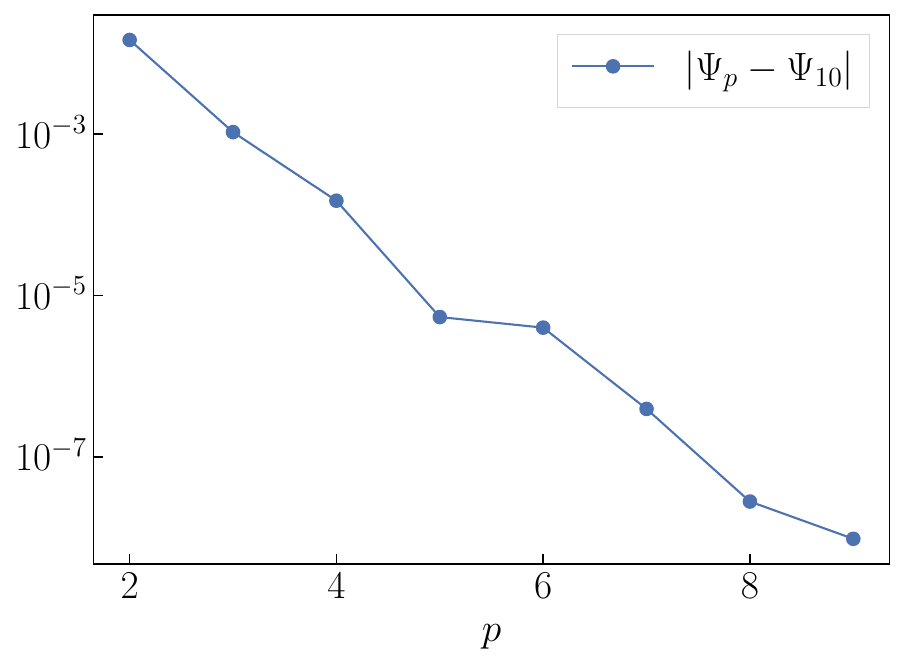}
    \caption{\textit{Convergence for binary BH system.} The system is the same as the right panel of Fig. \ref{fig: binary}. Here $|\Psi_p-\Psi_{10}|$ corresponds to a root-mean-square difference taken over 450 randomly selected points across the entire domain.
    }
    \label{fig: binary convergence}
\end{figure}

Figure~\ref{fig: binary convergence} demonstrates the numerical convergence of the solution with like charges. We compute solutions on computational domains where we vary the polynomial order $p$ in each element.  We interpolate each solution to a set of 450 randomly selected points across the entire domain, and compute the root-mean-square difference across these points between solutions at resolution $p$ with the highest resolution solution ($p=10$). The result is shown in Fig.~\ref{fig: binary convergence}, exhibiting exponential convergence of the scalar field profile for increasing resolution.

\begin{figure}
    \includegraphics[width = \linewidth]{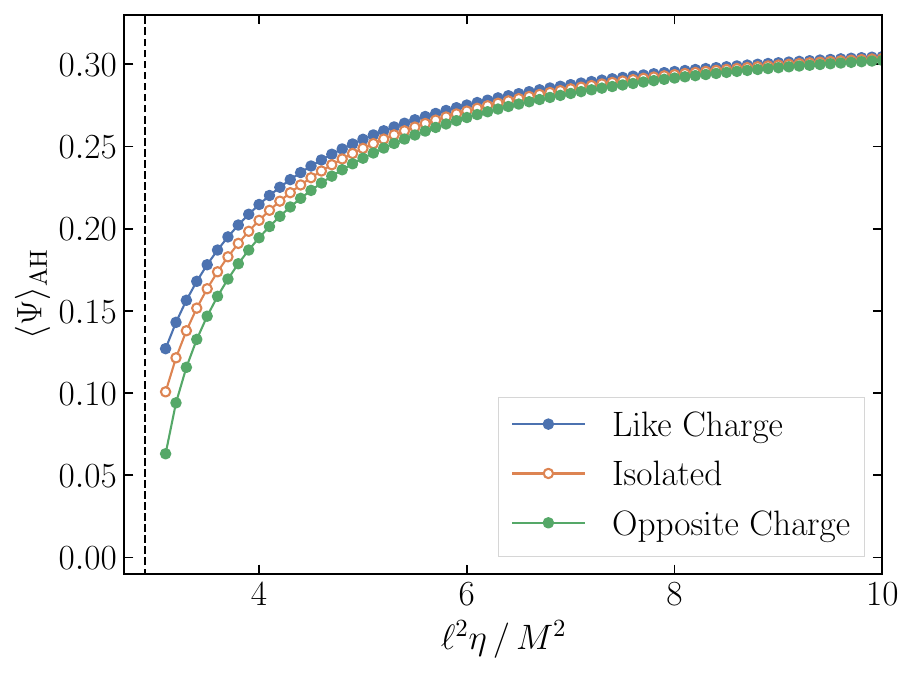}
    \caption{
        \textit{Relative difference of the scalar field between single BH and binary BH solves.} \(\langle\Psi\rangle_\mathrm{AH}\) indicates the average of the scalar field over the apparent horizon. The dashed line indicates the point at which, for single BH solves, no stable nonzero scalar hair profiles exist. The other BH is at a distance of $30.8M$, with both BHs initially at rest. We fix $\zeta=-10\eta$.
    }
    \label{fig: binary comparision}
\end{figure}

In a BH binary, the scalar hair near each BH is affected by the presence of the other. As a result of this interaction, the scalar configuration near each BH will differ from that of an isolated BH.
To quantify this effect, we calculate the average value of the scalar field \( \langle \Psi \rangle_{\mathrm{AH}}\) across one of the BH horizons.
Figure~\ref{fig: binary comparision} plots the value of \( \langle \Psi \rangle_{\mathrm{AH}}\) for an equal mass nonspinning BH binary, where the BHs are initially at rest,
for various values of the sGB coupling parameters.
For comparison, we also show \( \langle \Psi \rangle_{\mathrm{AH}}\) around a BH in isolation.
For larger couplings, we see that the influence of the opposite BH is smaller (typically a \(1\%\) difference). However, as we approach the existence threshold for scalarized solutions (dashed vertical line), the horizon average of the scalar field in the binary deviates further from that of an isolated BH.

\begin{figure}
    \includegraphics[width = \linewidth]{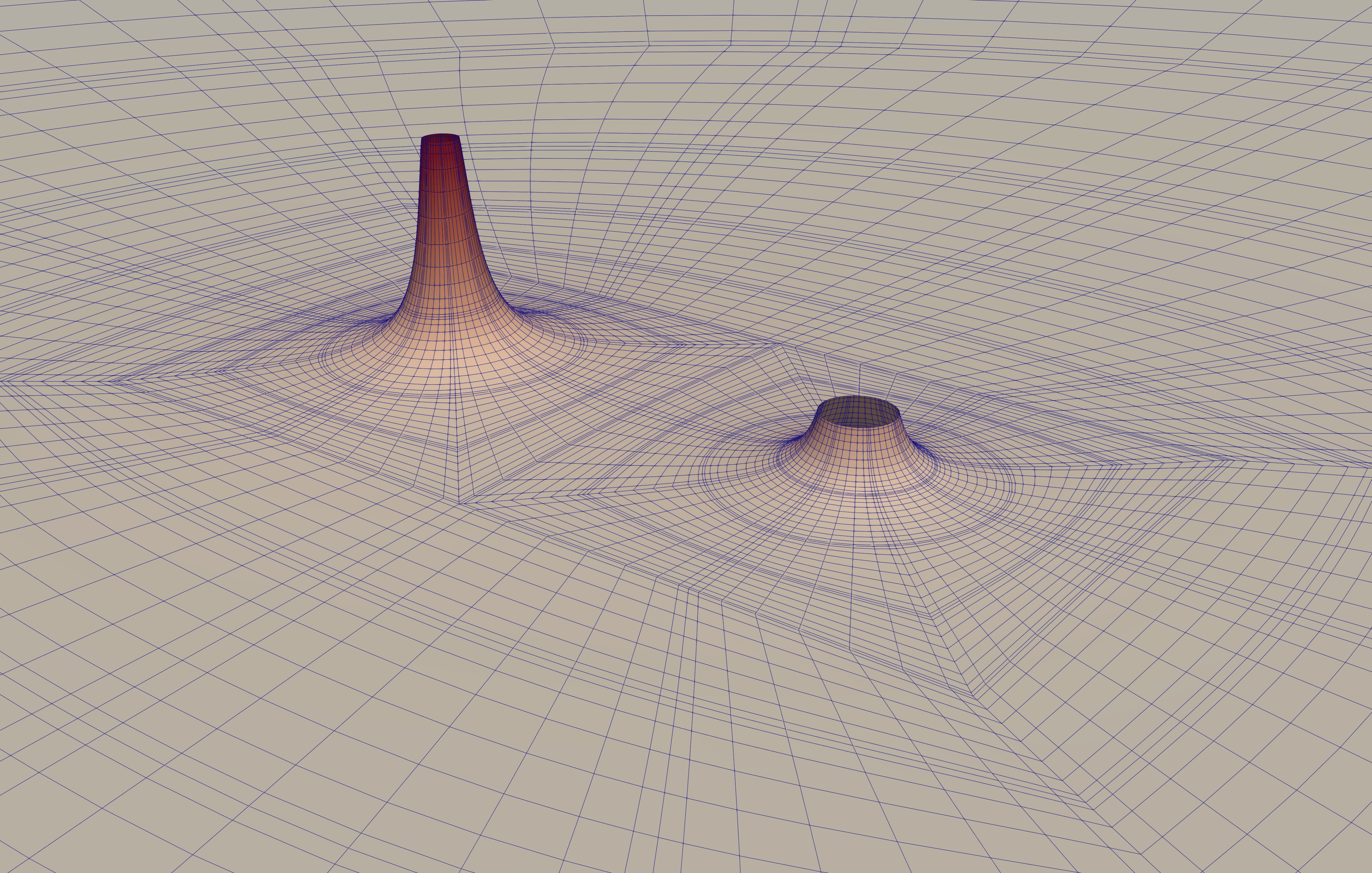} \\
    \vspace{0.5cm}
    \includegraphics[width = \linewidth]{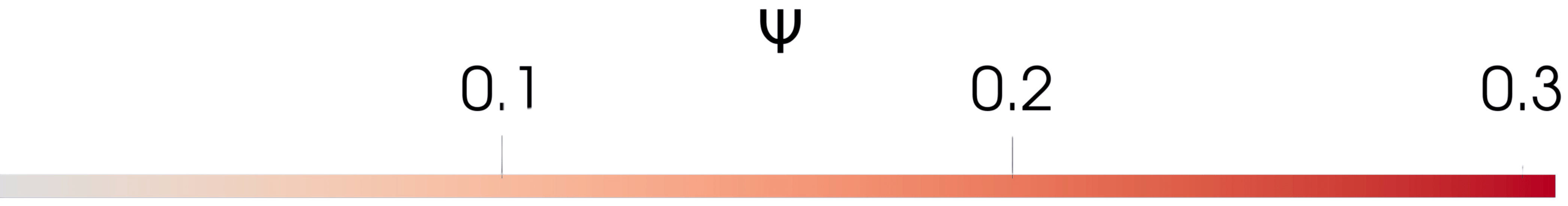}
    \caption{
        \textit{Scalar hair for a mass ratio 2 black hole binary.}
        Both black holes are nonspinning, with unequal mass $M_{1} = 2M_{2}$ and
        coupling constants \(\{\ell^2 \eta/M_1^2 = 2, \zeta=-10\eta\}\), at initial separation \(D / (M_1 + M_2) \simeq 15.4 \),
        in an approximately quasicircular configuration.
    }
    \label{fig: binary q2}
\end{figure}

Finally, moving toward more generic binary systems, Fig.~\ref{fig: binary q2} shows the scalar profile induced by a mass ratio 2 system. We use the same roll-off shape parameters as in Fig.~\ref{fig: binary}. If one were to consider both BHs as isolated/infinitely far away, only the smaller (left) BH would support a non zero stable scalar profile, whereas the larger BH (right) would not be scalarized.
However, the interaction between the two BHs leads to the larger BH also acquiring a scalar field.
Figures~\ref{fig: binary comparision} and~\ref{fig: binary q2} are a clear demonstration of scenarios where solving the augmented XCTS system (with the ``\(\partial_t = 0\)'' formulation) will lead to significantly different physics from the superposition of individual isolated solutions.


\section{Conclusion} \label{sec: conclusion}

This paper addresses the problem of constructing quasistationary initial data for black hole systems with scalar hair in scalar Gauss-Bonnet gravity.
We build upon the extended conformal thin sandwich approach in GR to propose a \textit{new} formulation in which quasistationary equilibrium of BH scalar hair is imposed.
The new system introduces an additional equation for the scalar field
obtained by requiring that the scalar gradient along the (approximate) time-like Killing vector of the spacetime vanishes.
The initial data obtained in this way represents an improvement with respect to the relaxation approach,
commonly used in the existing literature,
in which the scalar is allowed to develop (from an initial perturbation/guess) during the initial phase of time evolution.

We show that the additional scalar equation,
while being singular at black hole horizons,
is readily solvable with spectral methods.
We numerically implement the system
in the decoupling (test-field) limit
both in spherical symmetry, using a 1D \texttt{Python} code, as well as for generic spacetimes, using the elliptic solver~\cite{Vu:2021coj} in the open-source numerical relativity code \texttt{SpECTRE}~\cite{deppe_2024_11494680}.
As a comparison, we also implement the formulation of Kovacs~\cite{Kovacs:2021lgk}, and compare scalar profiles for single black hole spacetimes.
Through direct evolution we show that our new formulation indeed leads to stationary scalar hair, as opposed to scalar profiles constructed with the formulation of Ref.~\cite{Kovacs:2021lgk} that show initial transients.
Following this, we demonstrate that our 3D implementation performs robustly away from spherical symmetry, including boosted and/or rotating isolated black holes, as well as for binary black hole systems.

For binary systems, a further complication arises. Since the scalar solve is performed in the orbital \textit{comoving} frame, for which the coordinate velocities grow linearly with radius, there is a \textit{second} surface close to the light cylinder where the equations become singular.
We overcome this issue by deforming the equations with a \textit{roll-off} factor that regularizes the singular term in the far zone.
We show that the error introduced can approach truncation error near the black holes, while nearing $0.1\%$ in the far zone (where the scalar field is smaller).
It should be noted that, even for constraint-satisfying initial data in GR, evolutions typically take roughly one light-crossing time for the correct gravitational wave content to be present in the far-zone.
Since we expect the analogue of this to occur for the scalar radiation,
it is more important to ensure that near the black holes the system is as close to equilibrium as achievable
to reduce initial transients in the black holes parameters and trajectories.
Further, we have shown that, close to the scalar hair existence threshold, the quasistationary configuration for the binary is significantly affected by interaction of individual components--see Fig.~\ref{fig: binary comparision}.

While we have focused on scalar Gauss-Bonnet gravity, many technicalities encountered here will be common to other theories with additional scalar degrees of freedom, since quasistationarity of any additional fields can still be imposed with respect to the timelike Killing vector of the spacetime, and because the singular
behavior of the principal part
of the scalar equation is dictated solely by
the standard kinetic term, \(-\tfrac{1}{2} \nabla_{a} \Psi \nabla^{a} \Psi\), in the action.
For instance, singular behaviour of the principal part was found in the elliptic system specifying black hole initial data in damped harmonic gauge~\cite{Varma:2018evz}.
We also note that a formulation reminiscent of the one proposed here has been given in Ref.~\cite{Siemonsen:2023age} in the context of binary boson stars systems.
In that case, however, quasistationarity as it is imposed here cannot be imposed on the phase of the complex field, and no singular behaviour is expected close to the binary due to the lower compactness of boson stars.

While we have only implemented the new formulation in the decoupling limit, the next step is to allow the scalar field to backreact on the metric.
Even though this significantly alters the complexity of the equations, we believe that such modifications should introduce little additional technical difficulty. Specifically, given the effectiveness of the over-relaxation scheme for the scalar equation, the same approach will be taken in future work to solve the fully-coupled XCTS system.
It seems straightforward to treat the new interaction terms as fixed source terms during each relaxation iteration and, indeed, already a similar technique was applied in Ref.~\cite{Brady:2023dgu} to solve the metric sector of the constraint equations given a fixed scalar profile.

Our implementation already allows us to perform numerical relativity simulations with reduced transients and more precise control over the system being simulated. This opens up the possibility of more precise numerical experiments within this theory, as well as more detailed parameter space studies.


\begin{acknowledgments}
    The authors would like to thank Maxence Corman, Hector O.~Silva, Vijay Varma, and Nikolas A.~Wittek for fruitful discussions.
    Computations were performed on the Urania HPC systems at the Max Planck Computing and Data Facility.
    P.J.N. would like to thank many members of the ACR division for helpful feedback regarding a preprint of this paper.
    This work was supported in part by the Sherman Fairchild Foundation and by NSF Grants No.~PHY-2309211, No.~PHY-2309231, and No.~OAC-2209656.
\end{acknowledgments}


\bibliography{literature}

\end{document}